\documentclass[prd,twocolumn,pdflatex,showpacs,superscriptaddress]{revtex4-2}
\usepackage{graphicx}
\usepackage{dcolumn}
\usepackage{bm}
\usepackage{gensymb}
\usepackage{color}
\usepackage{subfig}
\usepackage{amssymb}
\usepackage{amsmath}
\usepackage{mathrsfs}

\DeclareMathAccent{\frmarr}{\mathord}{letters}{"7E}

\usepackage{color}

\usepackage{mathtools}
\begin{document}

\preprint{APS/123-QED}

\title{On the absence of shock waves and vacuum birefringence in Born--Infeld electrodynamics}

\author{Hedvika Kadlecov\'{a}}
\email{Hedvika.Kadlecova@eli-beams.eu}
\affiliation{Institute of Physics, Czech Academy of Sciences, Na Slovance 2, 18221, Prague, Czech Republic}

\date{\today}

\begin{abstract} 
We study the interaction of two counter--propagating electromagnetic waves in vacuum in the Born--Infeld electrodynamics. First we investigate the Born case for linearly polarized beams, ${\bf E}\cdot{\bf B}=0$, i. e. $\mathfrak{G}^2=0$ (crossed field configuration), which is identical for Born--Infeld and Born electrodynamics; subsequently we study the general Born--Infeld case for beams which are nonlinearly polarized, $\mathfrak{G}^2\neq0$. In both cases, we show that the nonlinear field equations decouple using self-similar solutions and investigate the shock wave formation. We show that the only nonlinear solutions are exceptional travelling wave solutions which propagate with constant speed and which do not turn into shocks for our approximation. We obtain two types of exceptional wave solutions, then we numerically analyze which phase velocities correspond to the counter-- or co--propagating beams and subsequently we determine the direction of propagation of the exceptional waves.
 

\end{abstract}

\maketitle


\section{\label{sec:Intro} Introduction}
The photon--photon scattering is one of the most important nonlinear processes in today's particle physics. The process breaks the linearity of the Maxwell equations and is one of the oldest predictions of quantum electrodynamics (QED). It occurs in vacuum
via the generation of virtual electron--positron pair creation resulting in vacuum polarization \cite{Electrodynamics}. 

The indirect measurement of this process was achieved only very recently in 2017. Off--shell photon--photon scattering \cite{KarplusNeuman} was indirectly observed in collisions of heavy ions accelerated by standard charged particle accelerators with $4.4\,\sigma$ significance. See review article \cite{Baur} and results of experiments obtained with the ATLAS detector at the Large Hadron Collider \cite{Aaboudetal,Aad2019} in 2017, where the cross--section was measured and identified as compatible with standard model QED predictions \cite{EnteriaSilveriaErratum,KlusekGawenda,Dremin,Klein}.

Such studies of fundamental physics become possible due to the increasing availability of high power lasers. This raises an interest in experimental observations and motivates theoretical studies of nonlinear QED in laser-laser scattering \cite{Mourou,Marklund,DiPizzaReview,BulanovEli, Tommasini,Parades,King,KogaBulanov,KarbsteinShaisultanov, GiesKarbsteinKohlfurstSeegert,BulanovCherenkov}, the interaction of relatively long wavelength radiation with X-ray photons \cite{Schlenvoigt,Shangai100PW,Heinzlliesfeld}, nonlinear laser--plasma interaction \cite{Shukla,Piazza1}, and complex problems on the boundary of nonlinear QED. 

Recently, we investigated creation of shock waves in the process of two counter--propagating waves in the QED quantum vacuum  \cite{KadlecovaKornBulanov2019}. We have used the Heisenberg--Euler approach in QED \cite{HeisenbergEuler,Electrodynamics,Gies} to investigate the process.  Later in \cite{KadlecovaMine2019}, we generalized our study. In detail, we analyzed the wave breaking direction of the electromagnetic wave which has backwards character for stronger fields. In the same setup of two counter--propagating electromagnetic waves,
we have investigated new relativistic electromagnetic soliton solutions and nonlinear waves in a quantum vacuum \cite{BulanovSoliton} and also high order harmonic generation \cite{HarmonicsSasBul2021}. Namely, the one-dimensional Korteveg-de-Vries (KdV) type solitons and the Kadomtsev-Petviashvily solitons ( multidimensional generalization of the KdV solutions). Solitons can propagate over a large distance without changing its shape which is important in experimental physics because such solitons can be measured.

In this paper, we want to concentrate on the process of two counter--propagating waves in the Born--Infeld electrodynamics and look at the shock wave creation which is more complicated than in QED vacuum because of the absence of birefringence.

The Born--Infeld electrodynamics \cite{BornInfeld1933, Born, BialynickiInfeld} represents a very unique nonlinear modification of the QED Lagrangian. Next to the development of the Heisenberg--Euler general expression for the quantum nonlinearities in the Lagrangian of QED  \cite{HeisenbergEuler,EulerKockel,KarplusNeuman}, there was an interest in a theory of QED with the upper limit on the strength of the electromagnetic field, later called Born--Infeld theory. Interestingly, the nonlinear process of photon--photon scattering is present in the Born--Infeld electrodynamics already at the classical level, such studies were conducted by Schr\"{o}dinger \cite{Schrodinger1, Schrodinger2}.

The Born--Infeld theory gained new interest in 1985, when it was found as a limiting case of string theory. In \cite{FradkinTseytlin}, it was found that the Born–-Infeld Lagrangian is the exact solution of a constant Abelian external vector field problem.
Besides the theoretical application to string theory, there is also an interest in experimental research in the Born--Infeld theory. Next to the search for photon--photon scattering in vacuum there is also need to test QED and non--standard models like Born--Infeld theory and scenarios where mini charged particles or axion--like bosons \cite{TommasiniFerrandoMichinelaSeco} are involved. The experimental observation and precision tests of the parameter for the Born--Infeld in the low energy effective Lagrangian are still aiming to reach the necessary sensitivity for its measurement in the process of photon--photon scattering. Newly, the PVLAS experiment \cite{CorrectPVLAS,Ruso2022,Fedotov2022,Karbstein2022,ReviewPVLAS2022} investigates measuring new limits also on the existence of hypothetical particles which couple to two photons, axion-like and milli-charged particles, besides casting upper limits on the magnetic birefringence predicted by QED.

The Born--Infeld electrodynamics behaves as an isotropic medium with a polarization--independent refractive index. The individual plane wave propagating at the speed of light in homogeneous isotropic scattering reduces its phase velocity uniformly \cite{RebhanTurk}. For example, in the process of photon--photon scattering, a counter--propagating, circularly polarized monochromatic wave of the same helicity serves as an isotropic medium for the other counter--propagating wave, studied from the classical perspective in \cite{Schrodinger2}.

As in the QED, in the limit of extremely intense electromagnetic fields, the Maxwell equations are modified due to the nonlinear process of photon-photon scattering that makes the vacuum refraction index depend on the field amplitude. Due to the nonlinearity of the field equations, the electromagnetic field interacts with itself and generates deformations in the light cone \cite{MeloMedeirosPompeia}.

The behaviour of shock waves in the Born--Infeld nonlinear electrodynamics has been studied thoroughly. An early theoretical analysis was made by Boillat who showed that both polarization modes travel along the light cone of one optical metric in exceptional nonlinear electrodynamics like Born--Infeld's \cite{BoillatI,BoillatII}. 
The term exceptional means that no shocks are formed \cite{BIGibbons,BoillatI} and that the fields on the wavefront always satisfy the field equations. Born-Infeld electrodynamics is called completely exceptional and it is the only completely exceptional regular nonlinear electrodynamics \cite{BialynickiInfeld}. The electrodynamics shows special features such as the absence of shock waves and birefringence. In \cite{BoillatIII}, the study was extended to the motion of more general discontinuity fronts, the discontinuities do not evolve into shocks, but when the shock exists at some initial time it propagates on characteristic surfaces, i.e. the Cauchy problem is well--posed. In fact, it was a conjecture presented in \cite{Brenier2004} that the shocks are not allowed to form in the Born-Infeld electrodynamics. A year later, the conjecture was shown as false in \cite{NevesSerre2005}, see also \cite{NevesSerreCauchy2005}. They show the appearance of shocks beyond the contact discontinuities in the resolution of the Riemann problem, at least when the initial-data are large. We will discuss our result with respect to these works in the discussion.


The general problem of shock wave development remains an open question. Quite recently there has been some progress in 3D, the general problem of shock formation was resolved by D. Christodoulou in \cite{Christodoulou2007} in 3D dimensions, where he proved that the shock waves are absent in 3D space for the Chaplygin gas known also as scalar Born--Infeld theory. 

In this paper, we investigate the problem of nonlinear wave evolution in a quantum vacuum in Born--Infeld electrodynamics. We are looking for a detailed theoretical description of the electromagnetic shock wave formation and its absence in the nonlinear quantum vacuum.  We present and analyze analytical solutions of the Born--Infeld electrodynamics field equations for the finite amplitude electromagnetic wave counter--propagating to the crossed electromagnetic field, i.e. two counter--propagating electromagnetic waves. Such configuration may correspond to the collision of a low--frequency, very high intensity laser pulse with a high frequency X--ray pulse generated by an XFEL. The first, long wavelength electromagnetic wave is approximated by a constant crossed field and the derived corresponding nonlinear field equations contain expressions for the relatively short wavelength pulse. The solutions of the nonlinear field equations are found in a form of the simple wave, also called the Riemann wave. We investigate the development of the shock waves, their formation and the wave steepening, in more detail. We show that the only nonlinear solutions of the Born--Infeld field equations in our approximation are nonlinear waves with constant phase velocities which do not turn into shocks, the so called exceptional travelling wave solutions. We discuss the absence of the shock formation process.

First, we have investigated the field equations of the Born Lagrangian for linearly polarized beams, which are identical to the equations for the Born--Infeld Lagrangian for the crossed field configuration ${\bf E}\cdot{\bf B}=0$, i.e. $\mathfrak{G}^2=0$. Second, we generalize the study to the more general case of nonlinearly polarized beams in the Born--Infeld Lagrangian: ${\bf E}\cdot{\bf B}\neq0$ and therefore $\mathfrak{G}^2\neq0$. In both cases, we show that the nonlinear field equations decouple using self-similar solutions and we investigate the shock wave formation. We show that the only nonlinear solutions in our approximation are exceptional travelling wave solutions which propagate with constant speed and which do not turn into shocks. In the Born case, we naturally obtain exceptional wave solutions for counter--propagating (real photon--photon scattering) and for a co--propagating (non-interacting) beam orientation we investigate their direction of propagation. In the Born--Infeld case, we have additionally chosen the solutions which have constant phase velocities to match the limits of phase velocities of the background field in the Born case. We obtain two types of exceptional wave solutions, we numerically analyze which phase velocities correspond to the counter-- or co--propagating beams and we determine their direction of propagation.

The paper is organized as follows: 
Section \ref{sec:Intro} serves as an introduction to our problem and we review the current state of knowledge about vacuum birefringence and absence of shock waves in Born--Infeld.

In Section \ref{sec:Lagrangian}, we review Born--Infeld and Born electrodynamics, and their field equations. 

In Section \ref{sec:DerivationEquations}, we derive the nonlinear field equations in Born theory, we add small amplitude perturbations and linearize the coefficients.

Specifically, the Section \ref{sec:DerivationEquations} is divided into: In Subsection \ref{sub:der}, we derive the Born field equations, in Subsection \ref{sub:derphase} we derive the phase velocity, in Subsection \ref{sub:lin}, we linearize the coefficients in the equations, in Subsection \ref{sec:SolvingEquations} we solve the Born field equations by assuming the solution in a form of a simple wave, we show that the system of equations decouple for the ordinary wave case. The solution has the form of a nonlinear wave without dispersion in the linear approximation. In Subsection \ref{sec:AnalyzingEquations}, we analyze the properties of the self-similar solutions. We analyze the solutions by the method of characteristics for two possible cases, $-$ and $+$, corresponding to two orientations of the beams, counter--propagating and co--propagating beams. We analyze the wave breaking and the character of the breaking wave for the $-$ and $+$ cases. We show that the only solutions are exceptional waves with constant phase velocities for our approximation.

In Section \ref{sec:AbsenceOfShockWavesBI}, we derive the field equations for our problem in Born--Infeld electrodynamics. 

Specifically:

In Subsection \ref{sec:Solve}, we solve the field equations for our problem in Born--Infeld electrodynamics. First, we add weak linear corrections to the fields and perform a linearization of the coefficients around the constant background field. We assume the solution in the form of a simple wave and show that the system of equations decouple. The solution has the form of a nonlinear wave without dispersion in the linear approximation. We also derive the phase velocities for the system of equations.

In Subsection \ref{sec:Disc}, we show that in the cases where the phase velocities are constant, the exceptional waves are the only solutions of the equations for our approximation and so we demonstrate the absence of shock waves in the Born--Infeld theory for the physically relevant solutions. We discuss solutions of the field equation of type I which are similar to the solutions in Born theory.
We analyze the solutions by the method of characteristics for two possible cases, $-$ and $+$. We discuss properties of the solutions, demonstrate that the shock wave steepening does not take place and that only the exceptional waves are created for our approximation.

In Subsection \ref{sub:secondEq}, we discuss solutions of the field equations of type II with non--zero right hand side where we choose the solutions for which the phase velocities are constant.

In Subsection \ref{sub:phaseVelPhys}, we plot the numerically calculated phase velocities to determine their value in order to see which case $-$ and $+$ occurs. According to this we are able to discuss the direction of propagation of the resulting nonlinear waves.

In Subsection \ref{sec:discussion} we summarize the properties of the solutions, their direction of movement and phase velocities for the two possible cases, $-$ and $+$. 


The main results of the paper are summarized in Section \ref{sec:conclusion}.
The Appendices \ref{sec:CoefRadiation}, \ref{sec:Coef66},  \ref{sec:Coef23}, \ref{sec:Coef32} and \ref{sec:Coef44} contain the detailed coefficients of the linearization performed in the paper.

\section{\label{sec:Lagrangian} Born--Infeld and Born electrodynamics}

\subsection{The Born--Infeld and Born Lagrangians}

The first model of a nonlinear electrodynamics was proposed by Born \cite{Born} in 1933 with the following choice of the Lagrangian,
\begin{equation}
\mathcal{L}_{B}=-b^2\left(\sqrt{1 - \frac{\bf{E}^2-\bf{B}^2}{b^2}}-1\right),\label{eq:LagrangianB}
\end{equation}
where ${\bf E}$ and ${\bf B}$ are electric and magnetic fields, $b$ being the free Born--Infeld constant (also known as the field strength parameter) having the dimension of the electromagnetic field and the units $c=\hbar=1$.
In more detail, the Born theory was described in \cite{BornInfeld1933}.

Born's motivation was to find classical solutions representing electrically charged particles with finite self-energy. The mechanism restricting the particle's velocity in relativistic mechanics to values smaller than $c$ is going to restrict the electric field in the Born theory with $\mathcal{L}_{B}$ (\ref{eq:LagrangianB}) to values smaller than the critical field $b$ (when ${\bf B}=0$) \cite{BialynickiInfeld}.

The Born theory was not satisfactory in several directions. The main difficulties were connected to the fact that the self--energy of a point charge is infinite.

A year later, the Born--Infeld electrodynamics was developed \cite{BornInfeld,BialynickiInfeld} with the Lagrangian given by
\begin{equation}
\mathcal{L}_{BI}=-b^2\left(\sqrt{1 - \frac{\bf{E}^2-\bf{B}^2}{b^2}-\frac{(\bf{E}\cdot\bf{B})^2}{b^4}}-1\right),\label{eq:LagrangianBI}
\end{equation}
where a new pseudoscalar invariant, the term $\mathcal{G}=\bf{E}\cdot\bf{B}$, was added to the Born--Infeld Lagrangian while maintaing the Lagrangian as relativistically covariant.

The Born and the Born-Infeld theories reduce to the linear Maxwell theory for fields which are much weaker than the critical field $b$, ($b\rightarrow \infty$, i.e., classical linear electrodynamics),
\begin{equation}
\mathcal{L}_{M}= \frac{1}{2}(\bf{E}^2-\bf{B}^2).\label{eq:Maxwell}
\end{equation}

The Born-Infeld theory is a unique nonlinear theory of the electromagnetic field because it is the only theory which does not lead to a birefringence effect, the propagation velocities in all directions do not depend on the wave polarization,  i.e. the velocity of light in the Born-Infeld theory does not depend on its polarization. The Maxwell theory and the nonlinear electrodynamics of Born and Infeld are the only relativistic theories in which this holds true, \cite{Bialynicka}.

Every theory of electrodynamic type is described by the source free Maxwell equations,
the first pair of Maxwell field equations reads,
\begin{align}
\nabla \cdot {\bf B}&=0,\nonumber\\
\nabla \times {\bf E}&=-{\partial_{t} {\bf B}}. \label{FirstMax2}
\end{align} 

The second pair can be found by varying the Lagrangian $\mathcal{L}_{BI}$ (\ref{eq:LagrangianBI}), which gives the field equations. The second pair of equations can be written as
\begin{align}
\nabla \times {\bf H}&=\partial_{t} {\bf D},\nonumber\\
\nabla \cdot {\bf D}&=0, \label{SecondMax2}
\end{align} 
together with the nonlinear constitutive relations,
\begin{equation}
{\bf D}=\frac{\partial{\mathcal{L}_{BI}}}{\partial{\bf E}},\; {\bf H}=-\frac{\partial{\mathcal{L}_{BI}}}{\partial{\bf B}}.\label{eq:CE}
\end{equation}


The consistency of the above equations and their relativistic covariance is guaranteed by the existence of a scalar Lagrangian density $\mathcal{L}_{BI}(\mathfrak{F},\mathfrak{G})$ \cite{Schwinger1951} in dependence on the so-called Poincar\'e invariants,
\begin{align}
\mathfrak{F}&=\frac{1}{4}F_{\mu \nu}F^{\mu \nu}=\frac{1}{2}\left({\bf B}^2-{\bf E}^2\right),\nonumber\\
\mathfrak{G}&=\frac{1}{4}F_{\mu \nu}\tilde F^{\mu \nu}={\bf E}\cdot{\bf B},\label{eq:mat2}\\
\tilde F^{\mu \nu}&=\frac{1}{2}\varepsilon^{\mu \nu \rho \sigma}F_{\rho \sigma},\nonumber
\end{align}
where $\varepsilon^{\mu \nu \rho \sigma}$ is the Levi-Civita symbol in four dimensions.

The equations (\ref{FirstMax2}) follow from the assumption of existence of potentials. Equations (\ref{SecondMax2}) follow from varying the Lagrange function $\mathcal{L}_{BI}(\mathfrak{F},\mathfrak{G})$.

The field equations for the Born--Infeld Lagrangian $\mathcal{L}_{BI}$ (\ref{eq:LagrangianBI}) are given by 
\begin{equation}
\partial_{\mu}\left(\frac{\partial\mathcal{L}_{BI}}{\partial(\partial_{\mu}{\Phi})}\right)-\frac{\partial{\mathcal{L}_{BI}}}{\partial\Phi}=0,\quad \Phi=(-\phi,\bf{A}).
\end{equation}


The Born (\ref{eq:LagrangianB}) and the Born--Infeld (\ref{eq:LagrangianBI}) Lagrangians can be rewritten in terms of Poincar\'e invariants as 
\begin{equation}
\mathcal{L}_{B}=-b^2\left(\sqrt{1+\frac{2\mathfrak{F}}{b^2}}-1\right),\label{eq:LagrangianBFG}
\end{equation}
and
\begin{equation}
\mathcal{L}_{BI}=-b^2\left(\sqrt{1+\frac{2\mathfrak{F}}{b^2}-\frac{\mathfrak{G}^2}{b^4}}-1\right).\label{eq:LagrangianBIFG}
\end{equation}

Born-Infeld electrodynamics is called completely exceptional and it is the only completely exceptional regular nonlinear electrodynamics \cite{BialynickiInfeld}. The electrodynamics shows special features as the absence of shock waves and birefringence.
However, as mentioned in the introduction, it was shown that the shocks appear beyond the discontinuities in the resolution of the Riemann problem, at least when the initial-data are large, \cite{NevesSerre2005}.

\section{\label{sec:DerivationEquations} Born field equations}
In this section, we will derive and analyze the field equations in the set up of two counter--propagating waves in vacuum in the Born electrodynamics. 

For the sake of brevity, we consider the two counter--propagating electromagnetic waves to be of the same polarization. We will work in the orthogonal coordinate system, $(x,y,z)$, where the two waves propagate along the $x-$axis. We assume the components of the waves as ${\bf E}=(0,0,E_{z})$ and ${\bf B}=(0,B_{y},0)$, which means that the term $\mathfrak{G}={\bf E}\cdot{\bf B}=0$. This is usually called a crossed field configuration. In this case the Born--Infeld Lagrangian (\ref{eq:LagrangianBI}) reduces to the Born Lagrangian (\ref{eq:LagrangianB}), hence the analysis can be done in the Born electrodynamics. 

In our setup, we investigate only the ordinary wave propagation from the birefringence effect. For studies, which include also the extraordinary wave and the nonlinear wave evolution in the full Born--Infeld electrodynamics, we will need to study nonlinearly polarized beams,  $\mathfrak{G}={\bf E}\cdot{\bf B}\neq0$, which will be investigated in the next section in Born--Infeld electrodynamics. 


The idea is to use our knowledge about solving the field equations in the Heisenberg--Euler approximation of the two counter--propagating electromagnetic waves to solve the field equations in the Born electrodynamics and subsequently in the Born--Infeld electrodynamics, see Section~\ref{sec:AbsenceOfShockWavesBI}.

\subsection{\label{sub:der} Derivation of Born field equations}
The field equations were found by varying the Lagrangian (\ref{eq:LagrangianB}) with respect to the potential ${\bf A}$ (the first set comes from the set of equations (\ref{FirstMax2}) and the second set from equations (\ref{SecondMax2})):
\begin{equation}
\partial_{t}B_{y}-\partial_{x}E_{z}=0,\label{eq:nonlinear1}
\end{equation}
\begin{align}
-&\left[1+\frac{E^2_{z}}{b^2}\frac{1}{\left(1-(E^2_{z}-B^2_{y})/b^2\right)}\right]\partial_{t}E_{z}\nonumber\\
+&\left[1-\frac{B^2_{y}}{b^2}\frac{1}{\left(1-(E^2_{z}-B^2_{y})/b^2\right)}\right]\partial_{x}B_{y}\nonumber\\
+&\frac{1}{b^2}\frac{E_{z}B_{y}}{\left(1-(E^2_{z}-B^2_{y})/b^2\right)}(\partial_{t}B_{y}+\partial_{x}E_{z})=0,\label{eq:nonlinearr}
\end{align}
where we denote $E_{z}\equiv E$ and $B_{y}\equiv B$ and the condition $1-1/b^2(E^2-B^2) > 0$ should be valid.

Subsequently, we add the small amplitude perturbation to the fields,
\begin{align}
E&=E_{0}+a_{z}(x,t),\nonumber\\
B&=B_{0}+b_{y}(x,t),\label{eq:EB1}
\end{align}
where the fields $E_{0}, B_{0}$ represent the constant electromagnetic background field and  $a_{z}(x,t)$, $b_{y}(x,t)$ are perturbations.  The equations (\ref{eq:nonlinear1}, \ref{eq:nonlinearr}) can be rewritten (using the expressions (\ref{eq:EB1})) in the following form:
\begin{align}
\partial_{t}b_{y}(x,t)&=\partial_{x}a_{z}(x,t), \label{eq:abequations}\\
\alpha\,\partial_{t}a_{z}(x,t)&-\beta\,[\partial_{x}a_{z}(x,t)+\partial_{t}b_{y}(x,t)]-\gamma\,\partial_{x}b_{y}(x,t)=0,\label{eq:shift}
\end{align}
where the coefficients $\alpha, \beta$ and $\gamma$ become,
\begin{align}
\alpha&=1+\frac{(E_{0}+a_{z})^2}{b^2}\frac{1}{a_{1}},\nonumber\\
\beta&=\frac{1}{b^2}\frac{(E_{0}+a_{z})(B_{0}+b_{y})}{a_{1}},\label{eq:ABC}\\
\gamma&=1-\frac{(B_{0}+b_{y})^2}{b^2}\frac{1}{a_{1}}\nonumber,
\end{align}
where $a_{1}=1-\cfrac{1}{b^2}\left[(E_{0}+a_{z})^2-(B_{0}+b_{y})^2\right]$.

\subsection{\label{sub:derphase} Derivation of the phase velocity}
Here we derive the coefficients of the background field to calculate the phase velocities. We assume that $a_{z}(x,t)=b_{y}(x,t)=0$ and obtain from Eqs.~(\ref{eq:ABC}) that
\begin{align}
\alpha_{0}&=\frac{1+\dfrac{B^2_{0}}{b^2}}{a_{2}},\;
\beta_{0}=\frac{E_{0}B_{0}}{b^2}\frac{1}{a_{2}},\;
\gamma_{0}=\frac{1-\dfrac{E^2_{0}}{b^2}}{a_{2}},\label{eq:ABCcrossedBI}
\end{align}
where $a_{2}=1-\cfrac{1}{b^2}(E^2_{0}-B^2_{0})$.

Furthermore, for the crossed field case, we choose $B_{0}=E_{0}$ for simplicity, we obtain
\begin{equation}
\alpha_{0}=1+\frac{E^2_{0}}{b^2},\quad
\beta_{0}=\frac{E^2_{0}}{b^2},\quad
\gamma_{0}=1-\frac{E^2_{0}}{b^2}.\label{eq:ABCcrossed}
\end{equation}

In order to find the wave phase velocity from the linearized equations, (\ref{eq:abequations}) and (\ref{eq:shift}), we look for solutions with the form:
\begin{equation}
a_{z} \propto \exp(-i\omega t + i qx),\;\; b_{y} \propto \exp(-i\omega t + i qx), \label{eq:ab}
\end{equation}
where $q$ is the wave number and $\omega$ is the frequency.
Substituting (\ref{eq:ab}) into the field equations (\ref{eq:abequations}, \ref{eq:shift}) for the background field with $\alpha=\alpha_{0}, \beta=\beta_{0}$ and $\gamma=\gamma_{0}$, (\ref{eq:ABCcrossedBI}), we obtain an algebraic set of equations for the wave velocity $v={\omega}/q$. 
Since the Born--Infeld medium is dispersionless,  the phase velocity, $v_{ph}=\omega/q$, and the group velocity, $v_{g}=\partial{\omega}/\partial{q}$, are equal: $v=v_{ph}=v_{g}$. 

Then we obtain the set of equations,
\begin{align}
a_{z}+vb_{y}&=0,\nonumber\\
v(b_{y}\beta_{0}-a_{z}\alpha_{0})-(a_{z}\beta_{0}+b_{y}\gamma_{0})&=0, \label{eq:vphasetwo}
\end{align}
which has two solutions,
\begin{align}
v_{1,2}&=\frac{-\beta_{0}\pm \sqrt{\beta^2_{0}+\alpha_{0}\gamma_{0}}}{\alpha_{0}}.\label{eq:vphsolutionExtra}
\end{align}

The expression under the square root can be simplified (using (\ref{eq:ABCcrossedBI})) as 
\begin{align}
\beta^2_{0}+\alpha_{0}\gamma_{0}=\cfrac{1}{a_{2}},\label{eq:vphsolutionExtraSqrt}
\end{align}
which results for the crossed field ($B_{0}=E_{0}$, (\ref{eq:ABCcrossed})) in 
\begin{equation}
\beta^2_{0}+\alpha_{0}\gamma_{0}=1. \label{eq:one}
\end{equation}

The velocities (\ref{eq:vphsolutionExtra}) can be simplified by using the expressions (\ref{eq:ABCcrossed}) and (\ref{eq:one}),
\begin{align}
v_{1,2}&=\frac{-\beta_{0}\pm 1}{\alpha_{0}}, \label{eq:vphasetwogen}
\end{align}
then we find the explicit velocities, 
\begin{align}
v_{1}&=-1,\label{eq:vphaseone}\\
v_{2}&=\frac{\gamma_{0}}{\alpha_{0}}=\frac{1-\cfrac{E^2_{0}}{b^2}}{1+\cfrac{E^2_{0}}{b^2}}. \label{eq:vphasetwo}
\end{align}

The phase velocities $v=v_{1,2}$ are the velocities for the wave propagation over the background crossed field in the Born theory. Let's us recall that the long wavelength pulse is being modelled as a constant field $E_{0}, B_{0}$ and the field equations contain expressions for the counter--propagating short wavelength pulse $a_{z}, b_{y}$. The solution $v_{1}$ corresponds to the co-propagating waves case $+$ and the solution $v_{2}$ corresponds to the case of counter--propagating waves case $-$ whose velocity is lower than speed of light $c$. 

The phase velocity also diminishes as the field strength parameter $b$ increases. In the limit $b\rightarrow \infty$, which leads to the linear Maxwell theory, the phase velocity $v_{2}\rightarrow 1$. The obtained result is used further as a limit case for the background crossed field.



In the following text, we will need to determine which solutions correspond to the two situations:
\begin{enumerate}
\item case $-$: the mutually interacting, counter--propagating waves in which the photon--photon scattering can happen.
\item case $+$: the co--propagating, non-interacting waves where photon--photon scattering does not occur.
\end{enumerate}

\subsection{\label{sub:lin} Linearization of the coefficients in the equations}
Now, we perform the linearization of the coefficients $\alpha, \beta$ and $\gamma$ about the constant background field,
\begin{align}
\alpha&=\alpha_{0}+\alpha_{a_{z}}a_{z}+\alpha_{b_{y}}b_{y},\nonumber\\
\beta&=\beta_{0}+\beta_{a_{z}}a_{z}+\beta_{b_{y}}b_{y},\label{eq:ABCdelta}\\
\gamma&=\gamma_{0}+\gamma_{a_{z}}a_{z}+\gamma_{b_{y}}b_{y},\nonumber
\end{align}
where we have denoted,
\begin{align}
\alpha_{a_{z}}&=(\partial_{a_{z}} {\alpha})|_{a_{z},b_{y}=0},\quad
\alpha_{b_{y}}=(\partial_{b_{y}}{\alpha})|_{a_{z},b_{y}=0},\nonumber\\
\beta_{a_{z}}&=(\partial_{a_{z}}{\beta})|_{a_{z},b_{y}=0},\quad
\beta_{b_{y}}=(\partial_{b_{y}}{\beta})|_{a_{z},b_{y}=0},\label{eq:alphabeta22}\\
\gamma_{a_{z}}&=(\partial_{a_{z}}{\gamma})|_{a_{z},b_{y}=0},\quad
\gamma_{b_{y}}=(\partial_{b_{y}}{\gamma})|_{a_{z},b_{y}=0}.\nonumber
\end{align}
Next, we need to expand the following expression
\begin{equation}
g(a_{z},b_{y})=\frac{1}{a_{1}},\label{eq:gexpand}
\end{equation}
into a Taylor series in two variables $a_{z},b_{y}$ around the point ($a_{z},b_{y}=0$), using this expansion, the parameters $\alpha$, $\beta$ and $\gamma$ become:
\begin{align}
\alpha&=1+\frac{(E_{0}+a_{z})^2}{b^2}g(a_{z},b_{y}),\nonumber\\
\beta&=\frac{1}{b^2}(E_{0}+a_{z})(B_{0}+b_{y})g(a_{z},b_{y}),\label{eq:ABCg}\\
\gamma&=1-\frac{(B_{0}+b_{y})^2}{b^2}g(a_{z},b_{y})\nonumber.
\end{align}

We perform the Taylor series for $B_{0}=E_{0}$, the crossed field configuration. The expansion then becomes
\begin{align}
g(a_{z},b_{y}) &\approx 1+\frac{2E_{0}}{b^2}(a_{z}-b_{y})\label{eq:gg}\\
+&\frac{1}{b^4}\left[(4E^2_{0}-b^2)a_{z}^2-8E^2_{0}a_{z}b_{y}+(4E^2_{0}+b^2)b^2_{y}\right]. \nonumber
\end{align}
In the following text, we use just the first two linear terms of (\ref{eq:gg}) in the linearization.
We identify the coefficients $\alpha_{a_{z}}, \beta_{a_{z}}, \gamma_{a_{z}}$ and $\alpha_{b_{y}}, \beta_{b_{y}}, \beta_{b_{y}}$ (\ref{eq:alphabeta22}), in the general formulae for $\alpha, \beta, \gamma$ (\ref{eq:ABC}), for the special choice, $B_{0}=E_{0}$, of the crossed field.


The coefficients (\ref{eq:alphabeta22}) have final form,
\begin{align}
\alpha_{a_{z}}&=\frac{2 E_{0}}{b^2}\left(1+\frac{E_{0}^2}{b^2}\right),\quad
\alpha_{b_{y}}=-2\frac{E^3_{0}}{b^4},\nonumber\\
\beta_{a_{z}}&=\frac{E_{0}}{b^2}\left(1+\frac{2E^2_{0}}{b^2}\right),\quad
\beta_{b_{y}}=\frac{E_{0}}{b^2}\left(1-\frac{2E^2_{0}}{b^2}\right),\label{eq:concrete}\\
\gamma_{a_{z}}&=-2\frac{E^3_{0}}{b^4},\quad
\gamma_{b_{y}}=\frac{2 E_{0}}{b^2}\left(\frac{E_{0}^2}{b^2}-1\right).\nonumber
\end{align} 

\subsection{\label{sec:SolvingEquations} Born self--similar solutions}

In this subsection, we solve the field equations for the Born electrodynamics in our approximation.

We approach solving the nonlinear equations using a Riemann wave (simple wave) which is well known in nonlinear wave theory \cite{KadomtsevKarpman, Kadomtsev, Whitham}. 
We have solved the field equations using the simple wave in the Heisenberg--Euler approximation in \cite{KadlecovaKornBulanov2019, KadlecovaMine2019}. Thanks to the similar structure of the field equations, we obtain similar solutions, but a difference comes from the Born Lagrangian (\ref{eq:LagrangianB}) in the form of different constant coefficients $\alpha_{a_{z}}, \alpha_{b_{y}}$, $\beta_{a_{z}}, \beta_{b_{y}}$ and $\gamma_{a_{z}}, \gamma_{b_{y}}$ (\ref{eq:concrete}).

We start with the field equations (\ref{eq:abequations}, \ref{eq:shift}) with parameter functions $\alpha(a_{z},b_{y}),  \beta(a_{z},b_{y})$ and $\gamma(a_{z},b_{y})$ (\ref{eq:ABC}) in the linear approximation (\ref{eq:ABCdelta}). 

\subsubsection{\label{sub:self} Self--similar solutions}
We are assuming the relation $b_{y}=b_{y}(a_{z})$, $\partial_{t}b_{y}=({\rm d} b_{y}/{\rm d} a_{z})\partial_{t}a_{z}$, and $\partial_{x}b_{y}=({\rm d} b_{y}/{\rm d} a_{z})\partial_{x}a_{z}$. The field equations ~(\ref{eq:abequations}, \ref{eq:shift}) become:
\begin{align}
\partial_{t}a_{z}&=\frac{{\rm d} a_{z}}{{\rm d} b_{y}} \partial_{x}a_{z},\label{eq:partOne}\\
\partial_{t}a_{z}&=\frac{1}{\alpha}\left(2\beta+\gamma\frac{{\rm d} b_{y}}{{\rm d} a_{z}}\right) \partial_{x}a_{z}.\label{eq:partTwo}
\end{align}

When we compare the two equations above, we get a quadratic equation for the function $b_{y}(a_{z})$ in the form
\begin{align}
\gamma\left(\frac{{\rm d}b_{y}}{{\rm d}a_{z}}\right)^2+2\beta\frac{{\rm d}b_{y}}{{\rm d}a_{z}}-\alpha=0,\label{eq:quadraticOrdinary}
\end{align}
which has two unique solutions
\begin{align}
\left(\frac{{\rm d}b_{y}}{{\rm d}a_{z}}\right)=\frac{-\beta\pm \sqrt{\beta^2+\alpha\gamma}}{\gamma}.\label{eq:solutionsI}
\end{align}

Furthermore, we assume the solution in the form
\begin{align}
\left(\frac{{\rm d}b_{y}}{{\rm d}a_{z}}\right)=\nu,\label{eq:linearSolutionsOO}
\end{align}
where we assume $\nu$ in the linearized form as
\begin{equation}
\nu=\nu_{0}+\nu_{a_{z}}a_{z}+\nu_{b_{y}}b_{y},\label{eq:nunu}
\end{equation}
with new parameters $\nu_{0}$, $\nu_{a_{z}}$ and $\nu_{b_{y}}$, which are derived later.
For the two solutions for $ {\rm d}b_{y}/{\rm d}a_{z}$ (\ref{eq:solutionsI}), we obtain two sets of parameters $\nu_{0}$, $\nu_{a_{z}}$ and $\nu_{b_{y}}$. We  discuss the two of them in the next subsection (\ref{sec:AnalyzingEquations}) where we investigate the wave steepening of the separate solutions.

Now, we need to derive general expressions for the parameters  $\nu_{0}$, $\nu_{a_{z}}$ and $\nu_{b_{y}}$. In the following calculation, we use the definition of tangent to a surface at a point $(\alpha_{0}, \beta_{0},\gamma_{0})$ as 
\begin{align}
f(\alpha,\beta,\gamma)&=f_{0}+f_{\alpha}(\alpha-\alpha_{0})+f_{\beta}(\beta-\beta_{0})\nonumber\\
&+f_{\gamma}(\gamma-\gamma_{0}),
\end{align}
where ${{\rm d}b_{y}}/{{\rm d}a_{z}}=f(\alpha,\beta,\gamma)$, and we have denoted
$f_{0}=f(\alpha,\beta,\gamma)|_{\alpha_{0},\beta_{0},\gamma_{0}},\,
f_{\alpha}=\partial_{\alpha}{f}|_{\alpha_{0},\beta_{0},\gamma_{0}},\,
f_{\beta}=\partial_{\beta}{f}|_{\alpha_{0},\beta_{0},\gamma_{0}},\,
f_{\gamma}=\partial_{\gamma}{f}|_{\alpha_{0},\beta_{0},\gamma_{0}}\nonumber,$
to indentify the parameters $\nu_{0}$, $\nu_{a_{z}}$ and $\nu_{b_{y}}$ in the form $\nu$ (\ref{eq:nunu}).

We obtain the resulting coefficients as
\begin{align}
\nu_{0}=&f_{0},\label{eq:nu0}
\end{align}
and
\begin{align}
f_{\alpha}&=\pm\frac{1}{2},\label{eq:nu1}\\
f_{\beta}&=\frac{1}{\gamma_{0}}\left(-1\pm\beta_{0}\right),\label{eq:nu2}\\
f_{\gamma}&=\pm\frac{\alpha_{0}}{2\gamma_{0}}-\frac{\left(-\beta_{0}\pm 1 \right)}{\gamma_{0}^2},\label{eq:nu3}
\end{align}
where we have used the expressions (\ref{eq:ABCdelta}) 
and the relation
\begin{equation}
\beta^2_{0}+\alpha_{0}\gamma_{0}=1.
\end{equation}

The linear coefficients $\nu_{0}$, $\nu_{a_{z}}$ and $\nu_{b_{y}}$ then have a final form,
\begin{align}
\nu_{0}=&f_{0},\nonumber\\
\nu_{a_{z}}=&\alpha_{a_{z}}f_{\alpha}+\beta_{a_{z}}f_{\beta}+\gamma_{a_{z}}f_{\gamma},\label{eq:nu}\\
\nu_{b_{y}}=&\alpha_{b_{y}}f_{\alpha}+\beta_{b_{y}}f_{\beta}+\gamma_{b_{y}}f_{\gamma}.\nonumber
\end{align}


The problem reduces to finding a solution to the differential equation (\ref{eq:linearSolutionsOO}). The equation has a form of total differential, therefore it can be solved by the method of integration factor, which we choose as $m(a_{z})=\exp(-\nu_{b_{y}}a_{z})$. 

The relation $b_{y}=b_{y}(a_{z})$, which solves the equation, has a structure,
\begin{equation}
\frac{1}{\nu_{b_{y}}}\exp{(-\nu_{b_{y}}a_{z})}\left((\nu_{0}+\nu_{b_{y}}b_{y})+\frac{\nu_{a_{z}}}{\nu_{b_{y}}}(\nu_{b_{y}}a_{z}+1)\right)=\delta,\label{eq:solutionImplicit}
\end{equation}
where $\delta$ is arbitrary constant. We can rewrite it and get the function $b_{y}=b_{y}(a_{z})$ explicitly:
\begin{equation}
b_{y}=\delta\,\exp(\nu_{b_{y}}a_{z})-\frac{\nu_{a_{z}}}{\nu_{b_{y}}}(\nu_{b_{y}}a_{z}+1)-\frac{\nu_{0}}{\nu_{b_{y}}}.\label{eq:solutionExplicit55}
\end{equation}

We determine the constant $\delta$ thanks to the initial condition $b_{y}|_{a_{z}=0}=0$,
\begin{equation}
\delta=\frac{\nu_{a_{z}}+\nu_{0}\nu_{b_{y}}}{\nu^2_{b_{y}}}.\label{eq:constOrdinary1}
\end{equation}

In order to use and stay in the linearized approximation, we perform Taylor expansion of the first term in (\ref{eq:solutionExplicit55}) to the first order, 
\begin{equation}
\exp{(\nu_{b_{y}}a_{z})}\approx 1+\nu_{b_{y}}a_{z}+\dots \label{eq:Taylor1}
\end{equation}
This produces the simplified first term in (\ref{eq:solutionExplicit55}) as
\begin{equation}
b_{y}=\delta\,(\nu_{b_{y}}a_{z}+1)-\frac{\nu_{a_{z}}}{\nu_{b_{y}}}(\nu_{b_{y}}a_{z}+1)-\frac{\nu_{0}}{\nu_{b_{y}}}.\label{eq:solutionExplicit1}
\end{equation}

In order to simplify the expression for $b_{y}$ (\ref{eq:solutionExplicit1}) even more, we substitute (\ref{eq:constOrdinary1}) into (\ref{eq:solutionExplicit1}), and obtain a solution which shows a linear relation between $a_{z}$ and $b_{y}$,
\begin{equation}
b_{y}=\nu_{0}a_{z}.\label{eq:baOrdinary}
\end{equation}

Let's get back to the field equations (\ref{eq:partOne}) and (\ref{eq:partTwo}) which we aim to solve. We rewrite equation ~(\ref{eq:partOne}) as
\begin{equation}\label{eq:finalOrdinary}
\partial_{t}a_{z}-\frac{1}{\nu}\partial_{x}a_{z}=0,
\end{equation}
where $\nu$ is given by equation (\ref{eq:nunu}).

In order to continue, we perform another linearization of the $1/\nu$ factor as 
\begin{align}
f(\nu)&=f(\nu)|_{\nu_{0}}+\partial_{\nu}{f}|_{\nu_{0}}(\nu-\nu_{0}),
\end{align}
and subsequently we obtain
\begin{align}
\frac{1}{\nu}=\frac{1}{\nu_{0}}\left(1-a_{z}\frac{\nu_{a_{z}}+\nu_{0}\nu_{b_{y}}}{\nu_{0}}\right).\label{eq:1nu}
\end{align}

\subsubsection{\label{sub:final} The final form of the nonlinear wave}
Using the previous results, we can write equation ~(\ref{eq:finalOrdinary}) with the factor $1/\nu$ (\ref{eq:1nu}) in the final form:
\begin{equation}\label{eq:finalOrdinary2}
\partial_{t}a_{z}+f(a_{z})\partial_{x}a_{z}=0,
\end{equation}
with the factor $f(a_{z})$ given by
\begin{equation}\label{eq:finalResult}
f(a_{z})=-\frac{1}{\nu_{0}}\left[1-a_{z}\frac{(\nu_{a_{z}}+\nu_{0}\nu_{b_{y}})}{\nu_{0}}\right].
\end{equation}

This is the final form of the equation which we use in the following analysis. In the limit $a_{z}=0$, the wave moves with the phase velocity of the unperturbed case $-1/\nu_{0}$. The solution contains the two possible solutions for ${\rm d}b_{y}/{\rm d}a_{z}$ (\ref{eq:solutionsI}), which are determined by two different sets of parameters $\nu_{0}$, $\nu_{a_{z}}$ and $\nu_{b_{y}}$.

In general, this form of equation contains the information whether the shock waves are being created, the wave steepening takes place and high--order harmonics are being generated. The two possible resulting wave equations have similar structure and the properties of the waves are hidden in the two sets of parameters $\nu_{0}$, $\nu_{a_{z}}$ and $\nu_{b_{y}}$ for the $+$ and $-$ solutions.  We will discuss the two branches of solutions in the next Subsection \ref{sec:AnalyzingEquations} where we investigate the wave steepening of the two possible solutions in more detail.

\subsection{\label{sec:AnalyzingEquations} Properties of Born self--similar solutions}
In this subsection we analyze the properties of equation (\ref{eq:finalOrdinary2}). The equation can be analyzed by the method of characteristics, we shortly review this method as well as wave breaking. Furthermore, we analyze the properties of the nonlinear electromagnetic wave in more detail.

\subsubsection{\label{sub:char} Method of characteristics and wave breaking}
We can solve the equation (\ref{eq:finalOrdinary2}) by the method of characteristics. The characteristic equations for the Eq.~(\ref{eq:finalOrdinary2}) are
\begin{equation}
\frac{{\rm d}x}{{\rm d}t}=f(a_{z}),\; \frac{{\rm d}a_{z}}{{\rm d}t}=0.
\end{equation} 
Their solutions are $a_{z}(x,t)=A_{0}(x_{0})$ and $x=f(A_{0}(x_{0}))t+x_{0}$, where the function $a_{z}(x,t)$ transfers along the characteristic $x_{0}$ without any distortion. Therefore for any differentiable function $A=A(x)$, we can write solution $a_{z}$ in a form
\begin{equation}
a_{z}(x,t)=A_{0}(x_{0})=A_{0}[x-f(a_{z}(x,t))t],\label{eq:AAaa}
\end{equation}
where $A_{0}$ is an arbitrary function determined by the initial condition, $a_{z}(x)|_{t=0}=A_{0}(x)$. 

Wave breaking is a typical behavior of waves in nonlinear dispersionless media. We can write the solution of equation (\ref{eq:finalOrdinary2}) in an implicit form (\ref{eq:AAaa}) with the Euler coordinate $x$ dependent on the Lagrange coordinate $x_{0}$ and time $t$.
The location where the wave breaks is determined by the gradient of function $a_{z}(x,t)$. The wave breaks when the gradient becomes infinite \cite{Panchenko}. We obtain such result by differentiating equation (\ref{eq:A}) as
\begin{align}
\partial_{x}a_{z}&=\frac{A'_{0}(x_{0})}{1+A'_{0}(x_{0})f'\,t}, \label{eq:gradient1}\\
\quad t_{br}&=-\frac{1}{A'_{0}(x_{0})f'},\label{eq:gradient}
\end{align}
where it is denoted
\begin{align}
A'(x_{0})&=\rm{d}A_{0}/\rm{d}x_{0},\\
f'&=\partial_{a_{z}}f(a_{z}).\label{eq:fdash}
\end{align}

The gradient becomes infinite at time $t_{br}$, when the denominator of equation (\ref{eq:gradient1}) vanishes at some point $x_{br}$. At the time $t_{br}$, when the wave breaks, the amplitude, $a_{z}(x_{br},t_{br})=a_{m}\sin{[k(x_{br}-f(a_{z}(x_{br},t_{br}))]}$, remains constant. Such singularity is called the wave breaking or the gradient catastrophe.

\subsubsection{\label{sub:char} The character of the breaking wave for the counter--propagating waves: the $-$ solutions in Born theory}
Here we will identify and concentrate on the $-$ solutions of equation ~(\ref{eq:solutionsI}). We identify the phase velocity as the phase velocity $v_{2}$ (see equation (\ref{eq:vphasetwo})), because the phase velocity decreases and becomes less than the speed of light $c$. We can also relate the parameter $\nu^{-}_{0}$ to the phase velocity $v_{2}$ by
\begin{equation}
\nu^{-}_{0}=-\frac{1}{v_{2}},\label{eq:newph}
\end{equation}
where $v_{2}>0$.

We can rewrite $f(a_{z})$ in equation (\ref{eq:finalOrdinary2}) by using the explicit expression for $\nu^{-}_{0}$ (\ref{eq:newph}):
\begin{equation}
f^{-}(a_{z})=v_{2}+a_{z}\frac{(\nu^{-}_{a_{z}}+\nu^{-}_{0}\nu^{-}_{b_{y}})}{{\nu^{-}_{0}}^2} \label{eq:finalResultfdash}.
\end{equation}

The final equation (\ref{eq:finalOrdinary2}) can be rewritten
in the standard form corresponding to the equation of nonlinear wave without dispersion \cite{KadomtsevKarpman,Kadomtsev}, 
\begin{equation}\label{eq:finalOrdinary22}
\partial_{t}\bar{a}_{z}+(v_{2}+\bar{a}_{z})\partial_{x}\bar{a}_{z}=0,
\end{equation}
where we have denoted
\begin{equation}
\bar{a}_{z}=\frac{(\nu^{-}_{a_{z}}+\nu^{-}_{0}\nu^{-}_{b_{y}})}{{\nu^{-}_{0}}^2} a_{z}.\label{eq:koef}
\end{equation}  

Therefore the direction of the wave breaking is given by the sign in front of the function $f'$ in equation (\ref{eq:fdash}). In order to investigate the wave steepening, we analyze the expression (\ref{eq:koef}) which we can rewrite using Eq. (\ref{eq:fdash}) as
\begin{equation}
\bar{a}_{z}=f' a_{z}, \quad f'=\frac{(\nu^{-}_{a_{z}}+\nu^{-}_{0}\nu^{-}_{b_{y}})}{{\nu^{-}_{0}}^2}.\label{eq:koefi}
\end{equation}

After performing the subtitution $\alpha_{0}$, $\beta_{0}$ and $\gamma_{0}$ (\ref{eq:ABCcrossed}), we observe that it is convenient to express the functions $f_{\alpha}, f_{\beta}, f_{\gamma}$ in terms of the phase velocity $v_{2}$. The explicit expressions are:
\begin{align}
f_{\alpha}&=- \frac{1}{2},\, f_{\beta}=-\frac{1}{v_{2}},\,f_{\gamma}=\frac{1}{2}\frac{1}{v_{2}^2}.\label{eq:ff11}
\end{align}

Then the coefficients $\nu_{a}, \nu_{b}$ (\ref{eq:nu}) become:
\begin{align}
\nu^{-}_{a_{z}}&=-\frac{E_{0}}{b^2v^2_{2}}\left[\left(\frac{1+E^2_{0}}{b^2}\right)v^2_{2}+\left(1+\frac{2E^2_{0}}{b^2}\right)v_{2}+\frac{E^2_{0}}{b^2}\right],\nonumber\\
\nu^{-}_{b_{y}}&=\frac{E_{0}}{b^2v^2_{2}}\left[\frac{E^2_{0}}{b^2}v^2_{2}-\left(1-\frac{2E^2_{0}}{b^2}\right)v_{2}+\left(\frac{E^2_{0}}{b^2}-1\right)\right]\label{eq:coefNew22}.
\end{align}

The function $f'^{-}$ ( Eq. (\ref{eq:finalResultfdash})) has the form
\begin{align}
f'^{-}=\frac{E_{0}}{b^2}&\left[-\left(1+\frac{E^2_{0}}{b^2}\right)v^2_{2}-\left(1+\frac{3E^2_{0}}{b^2}\right)v_{2}\right.\label{eq:f1res22}\\
&\left. +\left(1-\frac{3E^2_{0}}{b^2}\right)-\left(\frac{E^2_{0}}{b^2}-1\right)\frac{1}{v_{2}}\right].\nonumber
\end{align}

The steepening factor $f'^{-}$ in the general form (\ref{eq:f1res22}) is  expressed in terms of the phase velocity $v_{2}$ (\ref{eq:vphasetwo}) and the  the Born--Infeld constant $b$.  If a singularity is formed, the electromagnetic wave breaking creates a shock wave, which has a forward character for $f'^{-}>0$ and backwards character for $f'^{-}<0$. There is also a possibility that $f'=0$, then the shock waves are not created and only exceptional waves are the solutions of the equations \cite{BIGibbons,BoillatI}.

In the limit $b\rightarrow \infty$, which leads to the linear Maxwell theory, the steepening factor $f'^{-}\rightarrow 0$ and the phase velocity $v\rightarrow 1$. This corresponds to the fact that wave steepening does not happen in classical Maxwell theory. Subsequently, the resulting nonlinear wave equation (\ref{eq:koef}) with $f^{-}(a_{z})$ (\ref{eq:finalResultfdash}) becomes (in the limit to the Mawell theory):
\begin{equation}\label{eq:finalOrdinary3}
\partial_{t}a_{z}+v_{2}\partial_{x}a_{z}=0,
\end{equation}
where
\begin{equation}\label{eq:finalResult2}
f^{-}(a_{z})|_{b\rightarrow \infty}=v_{2}.
\end{equation} 

Continuing in the Born theory, after we substitute the phase velocity $v_{2}$ into equations (\ref{eq:coefNew22}) and (\ref{eq:f1res22}), we obtain the coefficients $\nu^{-}_{a_{z}}, \nu^{-}_{b_{y}}$ (\ref{eq:coefNew22}),
\begin{align}
\nu^{-}_{a_{z}}&=-\frac{2E_{0}}{b^2}\frac{\left(1+\cfrac{E^2_{0}}{b^2}\right)}{\left(1-\cfrac{E^2_{0}}{b^2}\right)},\quad
\nu^{-}_{b_{y}}=-\frac{2E_{0}}{b^2}\frac{1}{\left(1-\cfrac{E^2_{0}}{b^2}\right)},\nonumber
\end{align}
and importantly, the steepening factor becomes
\begin{equation}
f'^{-}=0.\label{eq:fdash-}
\end{equation}

This means that the only solutions for this case are exceptional waves. The exceptional travelling wave solutions propagate with constant speed and do not turn into shocks \cite{BIGibbons}.

Lastly, lets have a look at the shock wave steepening analytically. It does not take place, we can show it explicitly. The gradient (\ref{eq:gradient1}) and the time of wave breaking (\ref{eq:gradient}) are
\begin{align}
\partial_{x}a_{z}&=A'_{0}(x_{0}), \label{eq:gradient1+}\\
\quad t_{br}&=-\infty.\label{eq:gradient+}
\end{align}

The final form of the nonlinear wave equation for the case $-$ is
\begin{equation}\label{eq:finalOrdinary3+}
\partial_{t}a_{z}+v_{2}\partial_{x}a_{z}=0,
\end{equation}
and its solution (\ref{eq:A}),
\begin{equation}
a_{z}(x,t)=A_{0}(x_{0})=A_{0}[x-v_{2}t],\label{eq:A+}
\end{equation} 
propagates with constant phase velocity $v_{2}$ along the increasing $x$-axis.

This exceptional wave is the real contribution to the outgoing radiation from the photon--photon scattering in the Born electrodynamics.

\subsubsection{\label{sub:char} The character of the breaking wave for the co--propagating waves: the $+$ solutions in Born theory}

Here we will identify the $-$ solutions of equation ~(\ref{eq:solutionsI}). We identify the phase velocity of the resulting wave as $v_{1}=-1$.  Additionally, the parameter $\nu^{+}_{0}$ has the same value as the phase velocity $v_{1}$,
\begin{equation}
\nu^{+}_{0}=v_{1}=-1.\label{eq:newph1}
\end{equation}
We can rewrite the function $f(a_{z})$ (\ref{eq:finalResult}) as
\begin{equation}
f^{+}(a_{z})=f^{+}_{0}+a_{z}(x,t)f'^{+},\label{eq:fa}
\end{equation}
where 
\begin{equation}
f^{+}_{0}=-\frac{1}{\nu^{+}_{0}}=1,\quad f'^{+}=\frac{\nu^{+}_{a_{z}}+\nu^{+}_{0}\nu^{+}_{b_{y}}}{{\nu^{+}_{0}}^2}.\label{eq:ffff}
\end{equation} 

By substituting $\alpha_{0}$, $\beta_{0}$ and $\gamma_{0}$ (\ref{eq:ABCcrossed}) into $f_{\alpha}, f_{\beta}, f_{\gamma}$  (\ref{eq:nu}) for the case $+$, using again $v_{2}$, we obtain 
\begin{align}
f_{\alpha}&=\frac{1}{2},\, f_{\beta}=-1,\,f_{\gamma}=\frac{1}{2v_{2}}-\cfrac{1}{\left(1-\cfrac{E^2_{0}}{b^2}\right)}.\label{eq:ff11}
\end{align}

The coefficients $\nu^{+}_{a_{z}}, \nu^{+}_{b_{y}}$  (\ref{eq:nu}) become
\begin{align}
\nu^{+}_{a_{z}}&=\frac{E^3_{0}}{b^4v_{2}}\left[-1-v+\frac{2v_{2}}{\left(1-\cfrac{E^2_{0}}{b^2}\right)}\right],\nonumber\\
\nu^{+}_{b_{y}}&=\frac{E_{0}}{b^2v_{2}}\left[\left(\frac{E^2_{0}}{b^2}+1\right)v_{2}+\frac{E^2_{0}}{b^2}-1\right].\nonumber\\\label{eq:coefNew1}
\end{align}

The function $f'^{+}$ (\ref{eq:ffff}), expressed in terms of the phase velocity, has the form
\begin{align}
f'^{+}=\frac{E_{0}}{b^2}&\left[1-\frac{E^2_{0}}{b^2}+v_{2}\left(-1-\frac{2E^2_{0}}{b^2}+2\frac{E^2_{0}}{b^2}\frac{1}{\left(1-\cfrac{E^2_{0}}{b^2}\right)}\right)\right]\label{eq:reffff}.
\end{align}

The general form of the steepening factor $f'^{+}$ (\ref{eq:reffff}) is now expressed in the phase velocity $v_{2}$ (\ref{eq:vphasetwo}) and the $b$ is the Born--Infeld constant.

In the limit $b\rightarrow \infty$, which leads to the linear Maxwell theory, the steepening factor $f'^{+}\rightarrow 0$ and the phase velocity $v_{1}\rightarrow 1$. The resulting equation becomes 
\begin{equation}\label{eq:finalOrdinary4}
\partial_{t}a_{z}-\frac{1}{v_{1}}\partial_{x}a_{z}=0,
\end{equation}
where
\begin{equation}\label{eq:finalResult3}
f^{+}(a_{z})|_{b\rightarrow \infty}=-\frac{1}{v_{1}}.
\end{equation}

Now, lets continue the analysis in the Born theory. After using the phase velocity $v_{1}$, the coefficients (\ref{eq:coefNew1}) reduce to
\begin{align}
\nu^{+}_{a_{z}}&=0, \quad \nu^{+}_{b_{y}}=0,
\end{align}
and importantly the function (\ref{eq:reffff}):
\begin{equation}
f'^{+}=0.\label{eq:fdash+}
\end{equation}

In other words, the shock wave steepening does not take place in this case because the co--propagating waves do no interact and the photon--photon scattering does not occur. Only exceptional waves are created.

The final form of the nonlinear wave equation for the case $+$ has the form
\begin{equation}\label{eq:finalOrdinary3+}
\partial_{t}a_{z}+\partial_{x}a_{z}=0,
\end{equation}
and its solution,
\begin{equation}
a_{z}(x,t)=A_{0}(x_{0})=A_{0}[x - t],\label{eq:A+}
\end{equation}
which propagates to the right with the constant phase velocity $v_{1}=-1$ along the x-axis. The analytical expressions for the shock wave steepening are the same as for the previous case $-$, see equations (\ref{eq:gradient1+}) and (\ref{eq:gradient+}). There is no steepening for the exceptional waves.

\section{\label{sec:AbsenceOfShockWavesBI} Derivation of field equations in the Born--Infeld theory}

In this section we derive and analyze the field equations for the problem of two counter--propagating waves in Born--Infeld electrodynamics for 
nonlinearly polarized beams. In other words, we generalize our study from the previous section to include $\mathfrak{G}={\bf E}\cdot{\bf B}\neq 0$ in the calculations.
As mentioned in previous section, this generalized setup will give rise to extraordinary waves in theoretical nonlinear wave evolution in the full Born--Infeld electrodynamics.

We work in an orthogonal coordinate system, $(x,y,z)$, where the two waves propagate along the $x-$axis. We assume ${\bf E}=(0,E_{y},E_{z})$ and ${\bf B}=(0,B_{y},B_{z})$, which is the simplest generalization of the previous setup in the Born theory from Section \ref{sec:DerivationEquations}. The functions $E_{z}(t,x)$, $B_{y}(t,x)$ are functions of time $t$ and position $x$. However, the second equation in (\ref{FirstMax2}) allows us to assume an $E_{y}$ only dependent on $t$, and either $B_{z}$ dependent only on $x$ or $B_{z}$ equal to a constant. To simplify our ansatz we have chosen to use $B_{z}=0$ without loss of generality.


The field equations are found by varying the Lagrangian (\ref{eq:LagrangianBIFG}) according to the potential ${\bf A}$:
\begin{align}
\partial_{t}B_{y}&=\partial_{x}E_{z},\nonumber\\
-\alpha\partial_{t}E_{z} &+ \beta\left[\partial_{t}B_{y}+\tau\partial_{x}E_{z}\right]\nonumber\\
&+\gamma\partial_{x}B_{y}-\delta\partial_{t}E_{y}=0,\label{eq:nonl2}\\
-\epsilon\partial_{t}E_{z} &+ \zeta\partial_{x}B_{y}+\eta\partial_{t}B_{y}\nonumber\\
&+\theta\partial_{x}E_{z}-\iota\partial_{t}E_{y}=0,\nonumber
\end{align}

where the coefficients are: 
\begin{align}
\alpha&=1+\frac{E_{z}^2}{b^2}\frac{1}{S_{1}},\; \beta=\frac{1}{b^2}\frac{E_{z}B_{y}}{S_{1}},\; \delta=\frac{1}{b^2}\cfrac{E_{z}E_{y}}{S_{1}},\label{eq:coeffs}\\
\gamma&= 1-\cfrac{1}{b^2}E^2_{y}-\cfrac{B^2_{y}S_{4}^2+\cfrac{1}{b^2}E_{z}B_{y}E^2_{y}}{b^2 S_{1}}.\nonumber
\end{align}
The coefficients in the second set of field equations,
\begin{align}
\epsilon&=\frac{1}{b^2}E_{y}E_{z}\cfrac{S_{5}}{S_{1}},\;
\zeta=\frac{E_{y}E_{z}}{b^2}\left(1-\cfrac{B^2_{y}S_{2}}{b^2 S_{1}}\right),\label{eq:coeffs1}\\
\eta&= -\cfrac{E_{y}B_{y}}{b^2}\left(2-\cfrac{S_{2}S_{3}}{S_{1}}\right),\nonumber
\end{align}
and the coefficients in the third set of field equations,
\begin{align}
\theta&=\alpha \frac{E_{y}B_{y}}{b^2},\;
\iota= 1+\cfrac{1}{b^2}B^2_{y}+\cfrac{1}{b^2}E^2_{y}\cfrac{S_{2}S_{3}}{S_{1}},\;\tau=S_{4},\label{eq:coeffs2}
\end{align}
where
\begin{align}
S_{1}&=1 - \cfrac{\bf{E}^2-\bf{B}^2}{b^2}-\cfrac{(\bf{E}\cdot\bf{B})^2}{b^4},\nonumber\\
S_{2}&=1-\cfrac{1}{b^2}B^2_{y},\;
S_{3}=1+\cfrac{1}{b^2}E^2_{y},\label{eq:a2}\\
S_{4}&=1-\cfrac{1}{b^2}E^2_{y},\;
S_{5}=1+\cfrac{1}{b^2}B^2_{y}.\nonumber
\end{align}

We observe that the set of equations (\ref{eq:nonl2}) is the simplest generalization of our equations in Born electrodynamics (\ref{eq:nonlinear1}, \ref{eq:nonlinearr}).
The field equations (\ref{eq:nonl2}) reduce to the two field equations for Born theory (\ref{eq:nonlinear1}, \ref{eq:nonlinearr}) in Section \ref{sec:DerivationEquations} when we set $E_{y}=0$. The condition $S_{1} > 0$ holds. The equations describe both, the ordinary and the extraordinary wave propagation; the latter being determined by the only components $E_{y}(t)$ and $B_{z}=0$. 

\subsection{\label{sec:Solve} Solving the field equations}

\subsubsection{\label{sub:solvLinearizace} Adding weak linear corrections}
In this section, we add a small amplitude perturbation to the fields to solve the field equations. Then we perform a linearization procedure which also includes the linearization of the coefficients in the equations about the constant background field.

Again, we add the weak linear amplitude corrections to the fields,
\begin{align}
E_{y}&=E_{0}+a_{y}(t),\nonumber\\
E_{z}&=E_{0}+a_{z}(x,t),\label{eq:EB}\\
B_{y}&=B_{0}+b_{y}(x,t),\nonumber
\end{align}
where the fields $E_{0}, B_{0}$ represent the constant electromagnetic background field and  $a_{y}(t), a_{z}(x,t)$ and $b_{y}(x,t)$ are amplitude corrections.

After we substitute (\ref{eq:EB}) into the field equations~(\ref{eq:nonl2}),
these can be rewritten as
\begin{align}
\partial_{t}b_{y}(x,t)&=\partial_{x}a_{z}(x,t), \nonumber\\
-\alpha\,\partial_{t}a_{z}(x,t)&+\beta\,\left[\tau\partial_{x}a_{z}(x,t)+\partial_{t}b_{y}(x,t)\right]\nonumber\\
&+\gamma\,\partial_{x}b_{y}(x,t)-\delta\partial_{t}a_{y}=0,\label{eq:a2}\\
-\epsilon\,\partial_{t}a_{z}(x,t)&+\zeta\partial_{x}b_{y}(x,t)+\eta\partial_{t}b_{y}(x,t)\nonumber\\
&+\theta\,\partial_{x}a_{z}(x,t)-\iota\partial_{t}a_{y}=0.\nonumber
\end{align}

\subsubsection{\label{sub:solvLinearizace} Linearization of the coefficients}

We assume the linearized coefficients $\alpha, \beta, \gamma$,$\delta, \epsilon, \zeta$, $\eta, \theta, \iota, \tau$ about the constant background field in the form:
\begin{align}
\alpha&=\alpha_{0}+\alpha_{a_{z}}a_{z}+\alpha_{b_{y}}b_{y}+\alpha_{a_{y}}a_{y},\nonumber\\
\beta&=\beta_{0}+\beta_{a_{z}}a_{z}+\beta_{b_{y}}b_{y}+\beta_{a_{y}}a_{y},\nonumber\\
\gamma&=\gamma_{0}+\gamma_{a_{z}}a_{z}+\gamma_{b_{y}}b_{y}+\gamma_{a_{y}}a_{y},\label{eq:ABCdelta22}\\
 \vdots&\nonumber\\
\tau&=\tau_{0}+\tau_{a}a_{z}+\tau_{b}b_{y}+\tau_{y}a_{y},\nonumber
\end{align}

where we denote:
\begin{align}
\alpha_{a_{z}}&=(\partial_{a_{z}} {\alpha})|_{a_{z},b_{y},a_{y}=0},\nonumber\\
\alpha_{b_{y}}&=(\partial_{b_{y}}{\alpha})|_{a_{z},b_{y},a_{y}=0},\label{eq:alphabeta}\\
\alpha_{a_{y}}&=(\partial_{a_{y}}{\alpha})|_{a_{z},b_{y},a_{y}=0}\nonumber.
\end{align}
The other constant factors in the linearized coefficients (\ref{eq:ABCdelta22}) are similarly denoted. The coefficients are listed in their final forms in the Appendix~\ref{sec:Coef66} due to their lengthy character.

The constant parameters $\alpha_{0}, \beta_{0}, \gamma_{0}, \delta_{0}$, $\epsilon_{0}, \zeta_{0}, \eta_{0}, \theta_{0}, \iota_{0}$ and $\lambda_{0}$ are listed in Appendix~\ref{sec:CoefRadiation}. The parameters are obtained by setting $a_{z}(x,t)=b_{y}(x,t)=0$ in the linearized coefficients (\ref{eq:ABCdelta22}) and then simplifying them by setting $B_{0}=E_{0}$. This simplification was used in all the previous calculations to have compatible results. 

\subsubsection{\label{sub:phaseVel} The derivation of the phase velocity}

The phase velocity is derived from the linearized background equations (\ref{eq:a2}) by  using the same relations (\ref{eq:ab}) as in the Born model, since the medium is dispersionless in Born--Infeld. We denote the phase velocity by $v=v_{ph}=v_{g}$.

To obtain the algebraic expressions for $v$, we substitute expressions (\ref{eq:ab}) into the equations for the background field. We obtain the equations for the background field when we assume $a_{z}=b_{y}=a_{y}=0$ in the field equations~(\ref{eq:a2}), in other words, we set the coefficients $\alpha, \beta, \gamma, \dots $ to constant coefficients $\alpha_{0}, \beta_{0}, \gamma_{0}, \delta_{0}$, $\epsilon_{0}, \zeta_{0}, \eta_{0}, \theta_{0}, \iota_{0}$ and $\lambda_{0}$, where the coefficients are listed in Appendix~\ref{sec:CoefRadiation}.

We obtain the background field equations:
\begin{align}
a_{z}+vb_{y}&=0,\label{eq:n1}\\
-v(-\alpha_{0}a_{z}+\beta_{0}b_{y}-\delta_{0}a_{y})+\left[\gamma_{0}b_{y}+\beta_{0}\tau_{0}a_{z}\right]&=0, \label{eq:n2}\\
v(\epsilon_{0}a_{z}-b_{y}\eta_{0}+a_{y}\iota_{0})+(b_{y}\zeta_{0}+a_{z}\theta_{0})&=0.\label{eq:n3}
\end{align}

When substituting (\ref{eq:n1}) into (\ref{eq:n2}), we get a quadratic equation for the first phase velocity,
\begin{align}
-\alpha_{0}v^2-vM+\gamma=0,\label{eq:vpna2}
\end{align}
with solutions:
\begin{align}
v_{1,2}=\frac{M\pm\sqrt{M^2+4\alpha_{0}\gamma_{0}}}{-2\alpha_{0}},\label{eq:vph1}
\end{align}
where $M=\left(1+\tau_{0}\right)\beta_{0}-\cfrac{a_{y}}{b_{y}}\delta_{0}$. The solutions generalize those obtained with the Born theory (\ref{eq:vphasetwogen}). These velocities are properties of the ordinary wave. One velocity describes the real, physical, phase velocity for the case $-$, the other velocity corresponds to case $+$.

Similarly, when we substitute (\ref{eq:n1}) into (\ref{eq:n3}), we get a quadratic equation for the phase velocity,
\begin{align}
-\epsilon_{0}v^2-vY+\zeta_{0}=0,\label{eq:vpna21}
\end{align}
with two solutions:
\begin{align}
v_{3,4}=\frac{Y\pm\sqrt{Y^2+4\epsilon_{0}\zeta_{0}}}{-2\epsilon_{0}},\label{eq:vph2}
\end{align}
where $Y=\eta_{0}+\theta_{0}-\cfrac{a_{y}}{b_{y}}\iota_{0}$.
These phase velocities are new and we assume that they apply to the extraordinary wave. Only one of the velocities represents the physical one, i.e. corresponds to the case $-$. The other phase velocity corresponds to case $+$. We cannot derive more specific expressions now. In general, the phase velocities depend on the ratio of the two functions $a_{y}/b_{y}$ and the other coefficients are constant. We will investigate them more in detail later, see Subsection~\ref{sub:phaseVelPhys}.

\subsubsection{\label{sub:phaseVel} The simple wave solutions}

In this section we solve the equations by using the self--similar solutions that are well known in nonlinear wave theory \cite{KadomtsevKarpman, Kadomtsev, Whitham}.
As done previously, we use two related variables $b_{y}$ and $a_{z}$,  assuming $b_{y}=b_{y}(a_{z})$, but include a new function, $a_{y}(t)$, which is freely specified. Because of this term, the two beams have general nonlinear polarization, thus it is possible to investigate the birefringence effect.  

The analysis begins by taking the total differentials: $\partial_{t}b_{y}=({\rm d} b_{y}/{\rm d} a_{z})\partial_{t}a_{z}$, and $\partial_{x}b_{y}=({\rm d} b_{y}/{\rm d} a_{z})\partial_{x}a_{z}$. By using them in the linearized set of field equations (\ref{eq:a2}) we obtain the set of three equations:
\begin{align}
\partial_{t}a_{z}&=\frac{{\rm d} a_{z}}{{\rm d} b_{y}} \partial_{x}a_{z}, \label{eq:a11}\\
\partial_{t}a_{z}&=\frac{1}{\alpha}\left\{\partial_{x}a_{z}\left[(1+\tau)\beta+\frac{{\rm d}b_{y}}{{\rm d}a_{z}}\gamma\right]-\delta\partial_{t}a_{y}\right\},\label{eq:a22}\\
\partial_{t}a_{z}&\left(-\epsilon+\frac{{\rm d}b_{y}}{{\rm d}a_{z}}\eta\right)=-\partial_{x}a_{z}\left(\frac{{\rm d}b_{y}}{{\rm d}a_{z}}\zeta+\theta\right)+\iota\partial_{t}a_{y}.\label{eq:a33}
\end{align}

These can be solved by observing that equations ~(\ref{eq:a11}), (\ref{eq:a22}) and (\ref{eq:a33}) should be equal. This results in the set of two equations for $\partial_{x}a_{z}$ and $\partial_{t}a_{y}$:
\begin{align}
\partial_{x}a_{z}&\left\{1-\frac{1}{\alpha}\frac{{\rm d}b_{y}}{{\rm d}a_{z}}\left[(1+\tau)\beta+\frac{{\rm d}b_{y}}{{\rm d}a_{z}}\gamma\right]\right\}=-\partial_{t}a_{y}\frac{\delta}{\alpha}\frac{{\rm d}b_{y}}{{\rm d}a_{z}},
\label{eq:a221}\\
\partial_{x}a_{z}&\left[-\epsilon+\eta\frac{{\rm d}b_{y}}{{\rm d} a_{z}}+\frac{{\rm d}b_{y}}{{\rm d}a_{z}}\left(\zeta\frac{{\rm d}b_{y}}{{\rm d}a_{z}}+\theta\right)\right]=\iota\partial_{t}a_{y}\frac{{\rm d}b_{y}}{{\rm d}a_{z}}.\label{eq:a331}
\end{align}

The two equations above should be equal because they have the same form.  After substitution of one into the other, we obtain another quadratic equation for the total differential ${\rm d}b_{y}/{\rm d}a_{z}$:
\begin{align}
\left(\frac{{\rm d}b_{y}}{{\rm d}a_{z}}\right)^2&(\delta\zeta+\iota\gamma)+\left(\frac{{\rm d}b_{y}}{{\rm d}a_{z}}\right)\left[\eta\delta+\theta\delta-\iota\beta^2(1+\tau)\right]\nonumber\\
&+\alpha\left(\iota+\epsilon\frac{\delta}{\alpha}\right)=0. \label{eq:1234}
\end{align}

The quadratic equation (\ref{eq:1234}) has two solutions,
\begin{align}
&\left(\frac{{\rm d}b_{y}}{{\rm d}a_{z}}\right)_{1,2}=\frac{-N \pm \sqrt{N^2-4\alpha(\delta\zeta+\iota\gamma)(\iota-\epsilon(\delta/\alpha)}}{2(\delta\zeta+\iota\gamma)}, \label{eq:56}
\end{align}
where 
\begin{equation}
N=\left[ \delta(\eta+\theta)-\iota{\beta}^2(1+\tau)\right]. \label{eq:78}
\end{equation}

We have thus shown that the field equations decouple when we look for a solution in a simple wave form, which is one of the main results of this paper.

\subsection{\label{sec:Disc} Solutions of type I equations}
In this section we discuss the solutions of the decoupled field equations and their meaning. By type I equation we refer to an equation in a form (\ref{eq:a11}). This can be rewritten as
\begin{equation}
\partial_{t}a_{z}-\frac{1}{\nu}\partial_{x}a_{z}=0,\label{eq:firsteq}
\end{equation}
where, after linearization, the function $\nu$ has the form:
\begin{equation}
\frac{{\rm d}b_{y}}{{\rm d}a_{z}}=\nu,\quad \nu=\nu_{0}+\nu_{a_{z}}a_{z}+\nu_{b_{y}}b_{y}+\nu_{a_{y}}a_{y}.\label{eq:relation12}
\end{equation}

The type I equation has the same form as the equation (\ref{eq:finalOrdinary}) for Born  and the equation (50) for Heisenberg-Euler electrodynamics \cite{KadlecovaMine2019}, consequently the derivation below also generalizes the results in previous sections.

The two sets of coefficients $\nu_{0}$, $\nu_{a_{z}}$, $\nu_{b_{y}}$ can be derived for each of the total differentials $({\rm d}b_{y}/{\rm d}a_{z})_{1,2}$ (\ref{eq:56}) in Mathematica, using linearized coefficients (\ref{eq:ABCdelta22}) with the background coefficients in Appendix \ref{sec:CoefRadiation}. The resulting coefficients are listed in the Appendix~\ref{sec:Coef23} because of their complexity. 

Let us mention, for notation purposes, that we denote the two sets of coefficients as
\begin{equation}
\nu^{\pm}=\nu^{\pm}_{0}+\nu^{\pm}_{a_{z}}a_{z}+\nu^{\pm}_{b_{y}}b_{y}+\nu^{\pm}_{a_{y}}a_{y},\label{eq:nununu}
\end{equation}
where the  $\pm$ sign is motivated by the clear correspondence between the $-$ and $+$ and counter--propagating waves and co--propagating waves, respectively.

For simplicity, we use only the definition (\ref{eq:relation12}) in the following text. The $\pm$ notation is being used from the next subsection on.

We return to the solution of the differential equation (\ref{eq:relation12}). The equation can be rewritten as
\begin{equation}
\frac{{\rm d}b_{y}}{{\rm d}a_{z}}=\nu,\quad \nu=\nu'_{0}+\nu_{a_{z}}a_{z}+\nu_{b_{y}}b_{y},\label{eq:relation123}
\end{equation}
such that the terms on the right hand side are constant with respect to the variables in total differential. $\nu'_{0}$ is denoted by
\begin{equation}
\nu'_{0}=(\nu_{0}+\nu_{a_{y}}a_{y}),\label{nudash}
\end{equation} 
which is a constant with respect to variable $a_{z}$.

The equation can be solved by the method of integration factor, chosen as $m(a)=\exp(-\nu_{b_{y}}a_{z})$. The relation $b_{y}=b_{y}(a_{z})$ is determined by
\begin{equation}
\frac{1}{\nu_{b_{y}}}\exp{(-\nu_{b_{y}}a_{z})}\left((\nu'_{0}+\nu_{b_{y}}b_{y})+\frac{\nu_{a_{z}}}{\nu_{b_{y}}}(\nu_{b_{y}}a_{z}+1)\right)=\delta_{1},\label{eq:solutionImplicit}
\end{equation}
where $\delta_{1}$ is an arbitrary constant. Therefore the function $b_{y}=b_{y}(a_{z})$ has the form
\begin{equation}
b_{y}=\delta_{1}\,\exp(\nu_{b_{y}}a_{z})-\frac{\nu_{a_{z}}}{\nu_{b_{y}}}(\nu_{b_{y}}a_{z}+1)-\frac{\nu'_{0}}{\nu_{b_{y}}}.\label{eq:solutionExplicit}
\end{equation}

After Taylor-expanding the first term in equation (\ref{eq:solutionExplicit}), the constant $\delta_{1}$ can be determined by the initial condition $b_{y}|_{a_{z}=0}=0$:
\begin{equation}
\delta_{1}=\frac{\nu_{a_{z}}+\nu'_{0}\nu_{b_{y}}}{\nu^2_{b_y}}.\label{eq:constOrdinary}
\end{equation}

Equation (\ref{eq:constOrdinary}) can be used to express $b_{y}$:

\begin{equation}
b_{y}=\nu'_{0}a_{z}.
\end{equation}
This can be rewritten using $\nu'_{0}$ (\ref{nudash}) as
\begin{equation}
b_{y}=(\nu_{0}+\nu_{a_{y}}a_{y})a_{z},\label{eq:result15}
\end{equation}
which we need to linearize to
\begin{equation}
b_{y}=\nu_{0}a_{z}.\label{eq:result158}
\end{equation}
The function $b_{y}$ has the same form as in the Born (see Subsection \ref{sec:SolvingEquations}) and Heisenberg--Euler approximation \cite{KadlecovaKornBulanov2019,KadlecovaMine2019}.

After we substitute $b_{y}$ (\ref{eq:result15}) into $\nu$ (\ref{eq:relation123}), we obtain 
\begin{equation}
\nu=\nu_{0}+(\nu_{a_{z}}+\nu_{0}\nu_{b_{y}})a_{z}+\nu_{a_{y}}a_{y}+\nu_{b_{y}}\nu_{a_{y}}a_{y}a_{z},\label{eq:relation567}
\end{equation}
where we neglect the last nonlinear term due to linearization.
While solving equation (\ref{eq:firsteq}) we need to evaluate $1/\nu(a_{z},b_{y})$. We use a Taylor expansion in two variables which yields
\begin{equation}
\frac{1}{\nu}=\frac{1}{\nu_{0}}\left\{1-\frac{1}{\nu_{0}}\left[(\nu_{a_{z}}+\nu_{0}\nu_{b_{y}})a_{z}+\nu_{a_{y}}a_{y}\right]\right\}.\label{eq:1overnu}
\end{equation}

We rewrite equation~(\ref{eq:firsteq}) as
\begin{equation}
\partial_{t}a_{z}+f(a_{z},a_{y})\partial_{x}a_{z}=0,\label{eq:firsteqq22}
\end{equation}
where $f(a_{z},a_{y})=\cfrac{1}{\nu}$ can be summarized as
\begin{equation}
f(a_{z},a_{y})=-\frac{1}{\nu_{0}}\left\{1-\frac{1}{\nu_{0}}\left[(\nu_{a_{z}}+\nu_{0}\nu_{b_{y}})a_{z}+\nu_{a_{y}}a_{y}\right]\right\}.\label{eq:funkcevrovnici}
\end{equation}
Furthermore we can put equation~(\ref{eq:firsteqq22}) into a standard form \cite{KadomtsevKarpman, Kadomtsev} which describes a nonlinear wave without dispersion, 
\begin{equation}
\partial_{t}\overline{a}_{z}+\left(-\frac{1}{\nu_{0}}+\overline{a}_{z}+\overline{a}_{y}\right)\partial_{x}\overline{a}_{z}=0,\label{eq:firsteqqq}
\end{equation}
where  
\begin{equation}
\overline{a}_{z}=\frac{1}{\nu^2_{0}}(\nu_{a_{z}}+\nu_{0}\nu_{b_{y}})a_{z},\;
\overline{a}_{y}=\frac{\nu_{a_{y}}}{\nu^2_{0}}a_{y}.
\end{equation}
The final form of this equation contains information about the shock wave creation and subsequent effects such as higher-order harmonic generation. The formula is being solved for the variable $a_{z}$ while there is a free arbitrary function $a_{y}$. 

This nonlinear equation solves the type I equation and has a form similar to the nonlinear waves (\ref{eq:finalOrdinary2}) together with (\ref{eq:finalResult}) in Born, (see Subsection \ref{sec:SolvingEquations}) and the nonlinear waves (54) together with (55) in Heisenberg-Euler approximation \cite{KadlecovaKornBulanov2019,KadlecovaMine2019}, but with different constant coefficients.

In the limit $a_{z}=a_{y}=0$, the wave will move with the velocity $-1/\nu_{0}$, as in the unperturbed case. We have two solutions, $\nu_{0}=\nu^{\pm}_{0}$ ($\nu^{-}_{0}$ for counter--propagating waves and $\nu^{+}_{0}$ for co--propagating waves), see the summary section \ref{sec:discussion}. 

The above final equation is the general result with profile distortion by the presence of functions $a_{z}$ and $a_{y}$ which suggest that wave steepening takes place. But let's check the physical relevance of our results.

\subsubsection{\label{sub:secondEq} The characteristic equations}
We solve equation~(\ref{eq:firsteqq22}) by the method of characteristics. The characteristic equation for the equation~(\ref{eq:firsteq}) and the resulting equation for $a_{z}$ are
\begin{equation}
\frac{{\rm d}x}{{\rm d}t}=f(a_{z},a_{y}),\; \frac{{\rm d}a_{z}}{{\rm d}t}=0. \label{eq:chara}
\end{equation} 

For any differentiable function $A=A(x)$, we can write the self--similar solution $a_{z}$ as
\begin{equation}
a_{z}(x,t)=A_{0}(x_{0})=A_{0}[x-f(a_{z}(x,t), a_{y}(t))t],\label{eq:A}
\end{equation}
where $A_{0}$ is an arbitrary function determined by the initial condition $a_{z}(x)|_{t=0}=A_{0}(x)$.

\subsubsection{\label{sub:secondEqggg} The condition for existence of exceptional waves, wave steepening and physical solutions}

We use our knowledge about the phase velocities in Born electrodynamics (see Section~\ref{sec:DerivationEquations}): in the counter--propagating case $-$, the phase velocity $v_{2}$ (\ref{eq:vphasetwo})
is positive and less than the speed of light $c=1$, and photon--photon occurs. In the co--propagating case $+$, the phase velocity $v_{1}=-1$ (\ref{eq:vphaseone}); in this case the beams do not interact. We observe that the phase velocities are constant for both cases. In the general Born--Infeld electrodynamics we shall expect similar limits of the phase velocities. 
The phase velocities are dependent on the ratio $a_{y}/b_{y}$, $v_{1,2}=v_{1,2}(a_{y}/b_{y})$ and $v_{3,4}=v_{3,4}(a_{y}/b_{y})$,
where $a_{y}$ is an arbitrary function. 

In order to choose the relevant (physical) phase velocities for our problem of photon--photon scattering in Born--Infeld electrodynamics, we shall impose as a requirement the constant limiting values for the phase velocities obtained for Born electrodynamics, i.e. we should require the ratio $a_{y}/b_{y}$ to be a constant. This is also possible thanks to the behaviour in the Born--Infeld electrodynamics as an isotropic medium  with a polarization--independent refractive index.

In order to find the ratio, we start with expression (\ref{eq:result158}). This yields
\begin{equation}
\cfrac{a_{y}}{b_{y}}=\frac{a_{y}}{\nu_{0}a_{z}}=k_{BI},\label{eq:frac50}
\end{equation}
which tells us to look for the ratio $a_{y}/a_{z}$ and to determine the constant $k_{BI}$.

In order to find these, we start to study the wave steepening  (see Section (\ref{sub:char}) for a basic review, and  \cite{Whitham} for a general one). The wave steepening will happen forward for $\partial_{a_{z}}f(a_{z},a_{y}) > 0$, backwards for $\partial_{a_{z}}f(a_{z},a_{y}) < 0$, or we get only exceptional waves for $\partial_{a_{z}}f(a_{z},a_{y}) = 0$.

The characteristics (\ref{eq:chara}) have an envelope
\begin{align}
1&=-\partial_{a_{z}}f(a_{z}(x,t),a_{y}(t))t,\label{eq:envelope1}
\end{align}
from where we obtain
\begin{equation}
\partial_{a_{z}}f(a_{z},a_{y})=\frac{(\nu_{a_{z}}+\nu_{0}\nu_{b_{y}})+\nu_{a_{y}a}\nu_{0}k_{BI}}{\nu^2_{0}}, \label{eq:vexcept2}
\end{equation}
where we used $a_{y}=k_{BI}\nu_{0}a_{z}$ (\ref{eq:frac50}).

We can obtain the explicit expression for the constant $k_{BI}$ only for the exceptional wave:
\begin{equation}
k_{BI}=-\frac{(\nu_{a_{z}}+\nu_{0}\nu_{b_{y}})}{\nu_{a_{y}}\nu_{0}}.\label{eq:exceptiona1l}
\end{equation}
Using equation (\ref{eq:frac50}) we get
\begin{equation}
a_{y}=-\frac{(\nu_{a_{z}}+\nu_{0}\nu_{b_{y}})a_{z}}{\nu_{a_{y}}}\label{eq:exceptional}.
\end{equation}

In fact, the requirement of a constant phase velocity shows that the only possible waves which can satisfy this are the exceptional waves. Moreover it also determines the explicit form of the originally free function $a_{y}$ for all $x$.  


We get the final equation in the form:
\begin{equation}
\partial_{t}a_{z}+f(\nu_{0})\partial_{x}a_{z}=0,\label{eq:firsteqqQQ}
\end{equation}
subject to the initial condition $a_{z}(x)|_{t=0}=A_{0}(x)$, and where
\begin{equation}
f(\nu_{0})=-\frac{1}{\nu_{0}}.\label{eq:funkcevrovnici2Q}
\end{equation}
The self--similar solution is
\begin{equation}
a_{z}(x,t)=A_{0}(x_{0})=A_{0}\left[x+\frac{1}{\nu_{0}}t\right],\label{eq:AQ}
\end{equation}
where the velocity of propagation is the constant $-1/\nu_{0}$. The direction of motion for the two solutions $\nu_{0}=\nu^{\pm}_{0}$  is given by the sign of the velocities $\nu^{\pm}_{0}$. If $\nu^{\pm}_{0}>0$, the wave moves to the left, otherwise it moves to the right along the $x$ axis. $\nu^{-}_{0}$ refers to case $-$ and $\nu^{+}_{0}$ to case $+$ (see the results in the final summary \ref{sec:discussion} for detailed study of the direction of propagation).

To select the physical phase velocities for our problem required finding the constant ratio $a_{y}/a_{z}$. This is satisfied only for the exceptional waves in the solutions. Additionally, this sets the free function $a_{y}$ to a specific expression. Therefore we can claim that the first equation (\ref{eq:firsteq}) only has exceptional waves as physically relevant solutions for our case of photon--photon scattering process and that the wave steepening does not take place in this case.


\subsection{\label{sub:secondEq} Solutions of type II equations}

The other type of equation in our set are equations~(\ref{eq:a22}) and (\ref{eq:a33}), which we call type II equations. Here we choose the first one ~(\ref{eq:a22}) to investigate: 
\begin{equation}
\partial_{t}a_{z}+g(a_{z}, b_{y},a_{y})\partial_{x}a_{z}=-\frac{\delta}{\alpha}\partial_{t}a_{y},\label{eq:a222}
\end{equation}
where
\begin{equation}
g(a_{z},b_{y},a_{y})=-\frac{1}{\alpha}\left\{(1+\tau)\beta+\frac{{\rm d}b_{y}}{{\rm d}a_{z}}\gamma\right\}.
\end{equation}
We can rewrite this equation using the result for ${\rm d}b_{y}/{\rm d}a_{z}$ (\ref{eq:result158}) as 
\begin{align}\label{eq:gazay}
g(a_{z},b_{y},a_{y})&=-\frac{1}{\alpha}\left[(1+\tau)\beta+\gamma\nu\right].
\end{align}

The main difference from the previous equation is the non--zero right hand side which suggests the presence of a radiation source: a current determined by $\partial_{t}a_{y}$. Since $\partial_{t}a_{y}\neq 0$, $a_{z}$ will not be constant along the characteristics and in general, the characteristics will not be straight lines. In what follows, we investigate when the wave breaking might arise \cite{Whitham}.

Equation~(\ref{eq:a222}) can be reduced to an ordinary differential equation by the method of characteristics. This yields one characteristic equation,
\begin{equation}
\frac{{\rm d}x}{{\rm d}t}=g(a_{z}, b_{y}, a_{y}), \label{eq:fxt}
\end{equation}
and equation~(\ref{eq:a222}) reduces then to
\begin{equation}
\frac{{\rm d}a_{z}}{{\rm d}t}=-\frac{\delta}{\alpha}\partial_{t}a_{y}.\label{eq:dealpartiala}
\end{equation}

Before we proceed further, we need to linearize the function $g(a_{z},b_{y},a_{y})$ as
\begin{align}\label{eq:gazayzz}
g(a_{z}, b_{y}, a_{y})&=g_{0}+g_{a_{z}}a_{z}+g_{b_{y}}b_{y}+g_{a_{y}}a_{y},
\end{align}
where the explicit coefficients $g_{0}, g_{a_{z}}, g_{b_{y}}$ and $g_{a_{y}}$  can be found in Appendix \ref{sec:Coef32}. We linearize also the coefficient $\delta/\alpha$, denoting it as $q$,
\begin{align}\label{eq:qqq}
q=-\frac{\delta}{\alpha}=q_{0}+q_{a_{z}}a_{z}+q_{b_{y}}b_{y}+q_{a_{y}}a_{y},
\end{align}
where the coefficients $q_{0}, q_{a_{z}}, q_{b_{y}}$ and $q_{a_{y}}$ are listed in Appendix \ref{sec:Coef44}.

\subsubsection{\label{sub:consShock} The Cauchy initial condition}
Firstly, we analyze equation~(\ref{eq:a222}) without its right hand side. We can use the relation $b_{y}=\nu_{0}a_{z}$ (\ref{eq:result158}), obtaining 
\begin{align}\label{eq:gazayyy}
g(a_{z},a_{y})&=g_{0}+(g_{a_{z}}+g_{b_{y}}\nu_{0})a_{z}+g_{a_{y}}a_{y}.
\end{align}
The characteristic equations reduce to: 
\begin{equation}\label{eq:solucond}
\frac{{\rm d}x}{{\rm d}t}=g(a_{z},a_{y}),\; \frac{{\rm d}a_{z}}{{\rm d}t}=0,
\end{equation} 
while the self--similar solution for $a_{z}$ reads
\begin{equation}
a_{z}(x,t)=A_{0}(x_{0})=A_{0}[x-g(a_{z}(x,t), a_{y}(t))t],\label{eq:AA}
\end{equation}
where $A_{0}$ is an arbitrary function subject to the initial condition $a_{z}(x)|_{t=0}=A_{0}(x)$.

The envelope of characteristics becomes
\begin{equation}
1=-\partial_{a_{z}}g(a_{z},a_{y})t,\label{eq:envelope2}
\end{equation}
where 
\begin{equation}
\partial_{x_{0}}g(a_{z},a_{y})=(g_{a_{z}}+g_{b_{y}}\nu_{0})+g_{a_{y}}\nu_{0}k^{II}_{BI}.
\end{equation}
We have used the requirement for constant phase velocity (\ref{eq:frac50}), and a different constant $k^{II}_{BI}$ for the type II equation.
The wave breaks forward if $\partial_{a_{z}}g(a_{z},a_{y}) > 0$ or backwards if $\partial_{a_{z}}g(a_{z},a_{y}) < 0$; we get only exceptional waves if $\partial_{a_{z}}g(a_{z},a_{y}) = 0$.

We can determine the constant $k^{II}_{BI}$ only for the exceptional waves ($\partial_{a_{z}}g(a_{z},a_{y}) = 0$), then we obtain the explicit expression for $k^{II}_{BI}$ as
\begin{equation}
k^{II}_{BI}=-\frac{(g_{a_{z}}+g_{b_{y}}\nu_{0})+g_{a_{y}}\nu_{0}}{g_{a_{y}}\nu_{0}}.
\end{equation}
By using the expression (\ref{eq:frac50}) for $k^{II}_{BI}$, we obtain the final relation between $a_{z}$ and $a_{y}$ via the constant factor
\begin{equation}
a_{y}=-\frac{(g_{a_{z}}+g_{b_{y}}\nu_{0})}{g_{a_{y}}}a_{z}.\label{eq:exceptquadr}
\end{equation}

The choice of a constant phase velocity leads to exceptional waves as the only possibility. This also determines the explicit form of the originally free function $a_{y}$ for all x. As a consequence, wave steepening does not occur. 

We obtain the final equation in the form
\begin{equation}
\partial_{t}a_{z}+g_{0}\partial_{x}a_{z}=0,\label{eq:firsteqqQQq}
\end{equation}
where the explicit expression for $g_{0}$ (\ref{eq:g0}) is listed in Appendix D.

The self--similar solution for $a_{z}$ and for all $x$ reduces to
\begin{equation}
a_{z}(x,t)=A_{0}(x_{0})=A_{0}[x-g_{0}t],\label{eq:AQq}
\end{equation}
where the velocity of propagation is the constant $g_{0}$ moving along the $x-$axis. The direction of motion depends on the two solutions $g_{0}=g^{\pm}_{0}$. For the solutions $g^{\pm}_{0}>0$, the wave moves to the right and otherwise to the left along the $x-$axis, see the more detailed discussion in the final summary \ref{sec:discussion}.

As a result of selecting the physical phase velocities for our problem from all phase velocities for type II solutions, we observed the need to have the ratio $a_{y}/a_{z}$ as a constant. This condition is satisfied only for the exceptional waves in our solutions. Moreover, it sets the free function $a_{y}$ to a specific expression. 

\subsubsection{\label{sub:consShockRight} The solution with the right hand side}
The characteristic equation (\ref{eq:dealpartiala}) can be rewritten by substituting $b_{y}$ (\ref{eq:result158}) and $a_{y}$ (\ref{eq:exceptquadr}):
\begin{equation}
\frac{{\rm d}a_{z}}{{\rm d}t}=\left\{q_{0}+[q_{a_{z}}-g_{a_{z}}+\nu_{0}(q_{b_{y}}-g_{b_{y}})]a_{z}\right\}\partial_{t}a_{y}.\label{eq:qder}
\end{equation}
The characteristic equation is
\begin{equation}
\frac{{\rm d}a_{z}}{{\rm d}t}-Ma_{z}\partial_{t}a_{y}=q_{0}\partial_{t}a_{y},\label{eq:finalPartic}
\end{equation}
where 
\begin{equation}
M=q_{a_{z}}-g_{a_{z}}+\nu_{0}(q_{b_{y}}-g_{b_{y}}).
\end{equation} 

We solve the left hand side first and then proceed to find the particular solution for the right hand side. The solution of the equation
\begin{equation}
\frac{{\rm d}a^{0}_{z}}{{\rm d}t}=Ma_{z}\partial_{t}a_{y} \label{eq:finalLHS}
\end{equation}
is
\begin{equation}
a^{0}_{z}(x,t)=ce^{M[a_{y}(t)-a_{y}(0)]},\label{eq:azxyww}
\end{equation}
where $c \neq 0$ for all $x$.
To find the particular solution we integrate for $t > 0$, $t \in (0, t)$ and $-\infty < x < \infty$; obtaining
\begin{equation}
a^{p}_{z}(x,t)=-\frac{q_{0}}{M}.\label{eq:azxy}
\end{equation}

We obtain the general solution of equation~(\ref{eq:dealpartiala}) by combining equations (\ref{eq:azxyww}) and (\ref{eq:azxy}):
\begin{equation}
a^{c}_{z}(x,t)= -\frac{q_{0}}{M}+ce^{M[a_{y}(t)-a_{y}(0)]} \quad \label{eq:azxyzyyy}
\end{equation}
for all $x$ and the real constant $c$.

We assemble the final, singular solution by combining the solutions of equations~(\ref{eq:solucond}), the initial value solution (\ref{eq:AQq}) and the previous general solution. This results in
\begin{equation}
a^{f}_{z}(x,t)=A_{0}(x_{0})-\frac{q_{0}}{M}+ce^{M[a_{y}(t)-a_{y}(0)]}, \label{eq:azxyz}
\end{equation}
where we get $x=g_{0}t+x_{0}$ from the first characteristic equation (\ref{eq:fxt}).

The final solution was obtained by integrating the coupled ordinary differential equations~(\ref{eq:fxt}) and (\ref{eq:dealpartiala}). The initial value problem with data $a_{z}(x)=A_{0}(x)$ for $t=0$  is now modified by a constant $a^{f}_{z}(x)=A_{0}(x)-q_{0}/M$ for $t=0$.

Let's look explicitly at the function $g$ and the envelope of characteristics, respectively:
\begin{align}
g(a_{z},a_{y})&=g_{0}\nonumber\\
&+(g_{a_{z}}+g_{b_{y}}\nu_{0})[A_{0}(x_{0})-\frac{q_{0}}{M}+ce^{M[a_{y}(t)-a_{y}(0)]}]\nonumber\\
&+g_{a_{y}}a_{y} \label{eq:fin}
\end{align}
and  
\begin{equation}
1=-(g_{a_{z}}+g_{b_{y}}\nu_{0})\partial_{x_{0}}g(a_{z},a_{y})t.\label{eq:fin1}
\end{equation}
In the above, 
\begin{equation}\label{eq:fin2}
\partial_{x_{0}}g(a_{z},a_{y})=-\frac{1}{(g_{a_{z}}+g_{b_{y}}\nu_{0})}\partial_{x_{0}}A_{0}(x_{0}),
\end{equation}
and
\begin{equation}
\partial_{x_{0}}{A_{0}(x_{0})}=0,
\end{equation}
since $x=g_{0}t+x_{0}$ and $g_{0}$ is a constant. Therefore, we claim again that the type II equation~(\ref{eq:a222}) only has exceptional waves as physically relevant solutions and wave steepening does not occur.

The interpretation of these results is that the source term $a_{y}$, on the right hand side of equation~(\ref{eq:a222}), even in the form (\ref{eq:azxy}) is too weak to create a strong shock wave. Therefore the shock can not be produced \cite{Whitham}. 

To summarize, for the solutions of type I and type II equations we have restricted the phase velocities to physically consistent quantities. This allowed us to obtain the limit values for the photon--photon scattering process in Born electrodynamics. In the process we needed to find the constant ratio $a_{y}/a_{z}$. We have showed that such requirement is satisfied only by the exceptional waves in our solutions. This also determines the free function $a_{y}$ to a specific expression characteristic of each type of equations (I or II). 

In other words, the only physically relevant solutions to equations of type I or II are exceptional waves where wave steepening does not occur. 

\subsection{\label{sub:phaseVelPhys} The constant physical phase velocities}

We discuss our solutions with respect to the two cases of beam orientation, $-$ and $+$: we plot the numerical values of the phase velocities in Mathematica and look for their matching limiting value of the phase velocities, $v_{1}$ or $v_{2}$, for the background field in this section, see the summary in Section \ref{sec:discussion}.

It is enough to focus on the sign of the phase velocities. In other words, whether they approach a constant value with increasing $E_{0}$, i.e. one of the values which we have obtained for Born electrodynamics, $v_{1}$ (\ref{eq:vphaseone}) and $v_{2}$ (\ref{eq:vphasetwo}) (see Section~\ref{sec:DerivationEquations}) and which we have briefly reviewed in the beginning of Section \ref{sub:secondEqggg}.

\subsubsection{\label{sub:solv12} Identifying the phase velocities $v_{1,2}$}
The phase velocities originating from the first two equations in the set (\ref{eq:a2}) are the phase velocities $v_{1,2}$. We need to evaluate the ratio $a_{y}/b_{y}$ and plot $v_{1,2}=v_{1,2}(a_{y}/b_{y})$ as a function of $E_{0}$. 

Using the expression for exceptional waves (\ref{eq:exceptional}) and equation (\ref{eq:result158}), we obtain the ratio $a_{y}/b_{y}$ as
\begin{equation}
\cfrac{a_{y}}{b_{y}}=-\frac{\nu_{a_{z}}+\nu_{0}\nu_{b_{y}}}{\nu_{0}\nu_{a_{y}}},\label{eq:frac455}
\end{equation} 
which is a constant determined by two other possible sets of constants.

We start with the case $v_{1,2}$ and its coefficients  $\nu_{0}$, $\nu_{a_{z}}$, $\nu_{b_{y}}$ in order to visualize $v_{1,2}=v_{1,2}(a_{y}/b_{y})$ (\ref{eq:vph1}) and determine the two cases $+$ or $-$.
The expression for the velocities $v_{1,2}$ becomes constant using equation (\ref{eq:frac455}) as
\begin{align}
v_{1,2}=\frac{M\pm\sqrt{M^2+4\alpha_{0}\gamma_{0}}}{2\alpha_{0}},\label{eq:vph11}
\end{align}
where $M=(1+\tau_{0})\beta_{0}+\cfrac{\nu_{a_{z}}+\nu_{0}\nu_{b_{y}}}{\nu_{0}\nu_{a_{y}}}\delta_{0}$. 

Explicitly, using equation (\ref{eq:frac455}) and the two possible values of $\nu^{\pm}_{0}$ (\ref{eq:nupm}), $\nu^{\pm}_{a_{z}}$ (\ref{eq:nuazpm}), $\nu^{\pm}_{b_{y}}$ (\ref{eq:nubypm}) and $\nu^{\pm}_{a_{y}}$ (\ref{eq:nuaypm}), the velocities $v_{1,2}$ become: 
\begin{align}
v^{\pm}_{1}&=\frac{M+\sqrt{M^2+4\alpha_{0}\gamma_{0}}}{2\alpha_{0}},\nonumber\\
v^{\pm}_{2}&=\frac{M-\sqrt{M^2+4\alpha_{0}\gamma_{0}}}{2\alpha_{0}},\label{eq:vph111}
\end{align}
where $M=\left(1+\tau_{0}\right)\beta_{0}+\left(\cfrac{\nu_{a_{z}}+\nu_{0}\nu_{b_{y}}}{\nu_{0}\nu_{a_{y}}}\right)^{\pm}\delta_{0}$. 

\begin{figure}[!ht]
    \centering
   \includegraphics[width=0.9\columnwidth]{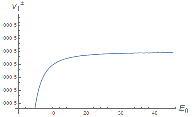}
    \caption{The phase velocities $v^{\pm}_{1}$. The phase velocities $v^{\pm}_{1}$ correspond to the counter--propagating case $-$, i.e. is positive ($v^{\pm}_{1}>0$), finite, and it appoaches a constant value.}\label{fig:phase1pm}
\end{figure}

\begin{figure}[!ht]    
    \centering
    \includegraphics[width=0.9\columnwidth]{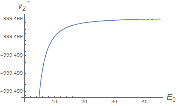}
    \caption{The phase velocities $v^{\pm}_{2}$. The phase velocities $v^{\pm}_{2}$ correspond to the co--propagating case $+$, i.e. is negative ($v^{\pm}_{2}<0$) and finite.}\label{fig:phase2pm}
\end{figure}

The phase velocities $v^{\pm}_{1}$ and $v^{\pm}_{2}$ are plotted in Fig.~\ref{fig:phase1pm} and Fig.~\ref{fig:phase2pm}. The phase velocities $v^{+}_{1}$ or $v^{-}_{1}$ correspond to the case $-$ because they approach the value of the phase velocity (\ref{eq:vphaseone}) which has a maximum of the speed of light $c=1$. Furthermore, $v^{+}_{2}$ or $v^{-}_{2}$ correspond to the case $+$ because they approach the value of the phase velocity (\ref{eq:vphasetwo}) which is $-1$. In the figures we use $E_{0}$ normalized to the Schwinger limit $E_{S}$ ($b=10^{-3}$) in order to see the positive and negative values of the phase velocities. The number for the parameter $b$ is chosen conveniently to demonstrate the phase velocities visually.



Even though we plot the different expressions for $v^{\pm}_{1}$, these have almost the same dependence on $E_{0}$ and furthermore, are positive. Therefore they correspond to the counter--propagating case $-$.

The near-identical behaviour of the overlaying curves in each figures~\ref{fig:phase1pm} and \ref{fig:phase2pm} could be attributed to the ratio $a_{y}/b_{y}$ which we discuss in the next paragraph. We observe that there are some numerical fluctuations as $E_{0}$ increases. These are consequences of the linear approximation of the coefficients that we performed in our calculation. 

\begin{figure}[!ht]
    \centering
\includegraphics[width=1.05\columnwidth]{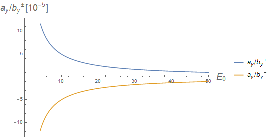}
    \caption{The ratio ${a_{y}/b_{y}}^{\pm}$ (\ref{eq:frac455}) is visualized as it depends on $E_{0}$.}\label{gr:newgraph1}
\end{figure}

Interestingly, if we visualize the ratio $a_{y}/b_{y}$ given by the constant expression (\ref{eq:frac455}) as it depends on $E_{0}$, we obtain Fig.~\ref{gr:newgraph1}. 


The ratio ${a_{y}/b_{y}}^{\pm}$ goes to zero very quickly as a function of $E_{0}$, both in negative values for ${a_{y}/b_{y}}^{-}$ and in positive values for ${a_{y}/b_{y}}^{+}$. Therefore, the contribution of this term is most relevant only early in the interaction. This also means that the source $a_{y}$ is small and insufficiently strong to more greatly influence the wave development.

\subsubsection{\label{sub:solv34} Identifying the phase velocities $v_{3,4}$}

We continue with the case $v_{3,4}$. These phase velocities originate from the first and the third equations in the set (\ref{eq:a2}). Below we present visualizations of $v_{3,4}=v_{3,4}(a_{y}/b_{y})$ (\ref{eq:vph2}) to determine which velocity corresponds to which case $+$ or $-$.

We obtain the ratio $a_{y}/b_{y}$ from the expression for exceptional waves (\ref{eq:exceptquadr}) and the relation for $b_{y}$ (\ref{eq:result158}),
\begin{equation}
\cfrac{a_{y}}{b_{y}}=-\cfrac{g_{a_{z}}+\nu_{0}g_{b_{y}}}{\nu_{0}g_{a_{y}}},\label{eq:frac457}
\end{equation} 
which is determined by the constants $g^{\pm}_{a_{z}}, g^{\pm}_{a_{y}}, g^{\pm}_{b_{y}}$ (\ref{eq:qazqbyqay}) and the two values of $\nu^{\pm}_{0}$ (\ref{eq:nupm}), $\nu^{\pm}_{a_{z}}$ (\ref{eq:nuazpm}), $\nu^{\pm}_{b_{y}}$ (\ref{eq:nubypm}), and $\nu^{\pm}_{a_{y}}$ (\ref{eq:nuaypm}).

The velocities $v_{3,4}$ become constant using equation (\ref{eq:frac457}). The phase velocities $v^{\pm}_{3}$ and $v^{\pm}_{4}$ can have two possible values thanks to two possible values of $\nu^{\pm}_{0}$ (\ref{eq:nupm}), $\nu^{\pm}_{a_{z}}$ (\ref{eq:nuazpm}), $\nu^{\pm}_{b_{y}}$ (\ref{eq:nubypm}) and $\nu^{\pm}_{a_{y}}$ (\ref{eq:nuaypm}).
The two possible values of $v^{\pm}_{3,4}$ can be expressed as
\begin{align}
v^{\pm}_{3}&=\frac{N +\sqrt{N^2+4\epsilon_{0}\zeta_{0}}}{-2\epsilon_{0}},\nonumber\\
v^{\pm}_{4}&=\frac{N - \sqrt{N^2+4\epsilon_{0}\zeta_{0}}}{-2\epsilon_{0}},\label{eq:vph3}
\end{align}
where $N=\eta_{0}+\theta_{0}+\left(\cfrac{g_{a_{z}}+\nu_{0}g_{b_{y}}}{\nu_{0}g_{a_{y}}}\right)^{\pm}\iota_{0}$.

\begin{figure}[!ht]
    \centering
    {\includegraphics[width=0.87\columnwidth]{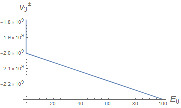}}
    \caption{The physical phase velocity $v^{\pm}_{3}$. The values are negative for both cases $v^{\pm}_{3}$, therefore the $v^{\pm}_{3}$ correspond to the co--propagating case $+$.}.\label{fig:phase3pm}
\end{figure}

\begin{figure}[!ht]
  \centering
    {\includegraphics[width=0.87\columnwidth]{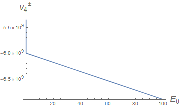}}
    \caption{The physical phase velocity $v^{\pm}_{4}$. The values are negative for both cases $v^{\pm}_{4}$, therefore the $v^{\pm}_{4}$ correspond to the co--propagating case $+$.}.\label{fig:phase4pm}  
\end{figure}

The phase velocities $v^{+}_{3}$ and $v^{-}_{4}$ are plotted in figures ~\ref{fig:phase3pm} and ~\ref{fig:phase4pm}. The phase velocities $v^{\pm}_{3}$ and $v^{\pm}_{4}$ seem to correspond to the co--propagating case $+$ since their values are negative. 

The near-identical behaviour of the overlaying curves in each figures~\ref{fig:phase3pm} and \ref{fig:phase4pm} could be attributed to the fraction $a_{y}/b_{y}$.  In this case, its contribution is negligible and the curve remains linear.

\subsection{\label{sec:discussion} Discussion of the validity of the linear approximation}
It is neccessary to discuss the previous results. 
Since the nonlinear Born--Infeld electrodynamics describes the electromagnetic fields in an isotropic medium with a polarization--independent refractive index \cite{RebhanTurk, Schrodinger2}, the extraordinary wave should move in the same direction as the ordinary wave with the same phase velocity.

According to our results, the phase velocities for the extraordinary wave give different phase velocities (\ref{eq:vph3}) than the ones for the ordinary wave (\ref{eq:vph111}). When looking at the plots of $v^{\pm}_{3}$ and $v^{\pm}_{4}$, the figures \ref{fig:phase3pm} and \ref{fig:phase4pm} show incomplete results with very low phase velocities only for the co--propagating case $+$. The physical phase velocities are missing. 
It is caused definitely by the insufficient precision of the linear approximation we used to study extraordinary wave in our paper. In order to obtain also the physical phase velocities and the same phase velocites $v^{\pm}_{1,2}=v^{\pm}_{3,4}$, we would have to use approximation to higher orders of expansion in the fields.
The linearization was good enough to study the ordinary wave, the phase velocity $v_{1,2}$, because it gives consistent results also for physical velocities. Using the higher order approximation will motivate and require further analysis.

In our study, the exceptional wave solutions in Born--Infeld electrodynamics do not turn into shocks in our approximation. Such result is in agreement with publications \cite{Boillat1972,BoillatI,BoillatII}, and the conjecture \cite{Brenier2004}, that the shocks are not allowed to form in the Born-Infeld electrodynamics. Since the conjecture was shown as false \cite{NevesSerre2005}, there is still possibility of their existence. It would be necessary to investigate which initial conditions are large enough to create the shock waves or use higher order approximation in the fields to see whether that would enable the shock wave creation.

\subsection{\label{sec:discussion} Summary  - Direction of propagation of the solutions}
In this section we summarize the main results and discuss the direction of propagation of the resulting waves, thanks to the investigation in section \ref{sub:phaseVelPhys} above.

\subsubsection{\label{sec:typeI} The type I solutions}
The type I equation (\ref{eq:firsteq}) has the form
\begin{equation}
\partial_{t}a_{z}+f(\nu_{0})\partial_{x}a_{z}=0,\label{eq:firsteqqQQ}
\end{equation}
and is subject to the initial condition $a_{z}(x)|_{t=0}=A_{0}(x)$. In equation (\ref{eq:firsteqqQQ}),
\begin{equation}
f(\nu_{0})=-\frac{1}{\nu_{0}}. \label{eq:funkcevrovnici2Q}
\end{equation}
The self--similar solution is given by
\begin{equation}
a_{z}(x,t)=A_{0}(x_{0})=A_{0}\left[x+\frac{1}{\nu_{0}}t\right],\label{eq:AQ}
\end{equation}
where the velocity of propagation is the constant $-1/\nu_{0}$.

\begin{figure}[!ht]
    \centering
  \includegraphics[width=1\columnwidth]{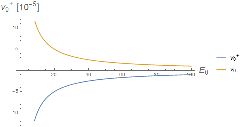}
    \caption{The coefficients $\nu^{\pm}_{0}$ visualized as a function of $E_{0}.$}\label{fig:coefnupm}
\end{figure}

The direction of motion is given by the two values for $\nu_{0}=\nu^{\pm}_{0}$ (\ref{eq:nupm}) which is finally visualized in Fig.~\ref{fig:coefnupm}, where we have normalized $E_{0}$ to the Schwinger limit $E_{S}$ and used $b=10^{-3}$. We observe that  $\nu^{-}_{0}>0$, therefore the wave moves to the right along the $x$ axis and corresponds to the counter--propagating case $-$ of the two beams. We observe that $\nu^{+}_{0}<0$, therefore the wave moves to the left along the $x$ axis and corresponds to the co--propagating case $+$ of the two beams. The results align with those in Born, see Section~\ref{sec:DerivationEquations}.


\subsubsection{\label{sec:typeII} The type II solutions}
\begin{figure}[!ht]
    \centering
\includegraphics[width=1\columnwidth]{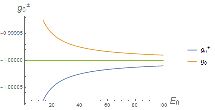}
    \caption{The coefficients $g^{\pm}_{0}$ as a function of on $E_{0}$. The horizontal line at the value $-1$ is there for comparison and represents the phase velocity limit for the case $+$ in Born electrodynamics.}\label{fig:g0pm}
\end{figure}

Type II of equations (\ref{eq:a222}) have the form
\begin{equation}
\partial_{t}a_{z}+g^{\pm}_{0}\partial_{x}a_{z}=q\partial_{t}a_{y},\label{eq:fin}
\end{equation}
where  the explicit expression for $g^{\pm}_{0}$ is given by equation (\ref{eq:g0}), and the final singular solution with $x=g^{\pm}_{0}t+x_{0}$ has the form
\begin{equation}
a^{f}_{z}(x,t)=A_{0}[x-g^{\pm}_{0}t]-\frac{q_{0}}{M}+ce^{M[a_{y}(t)-a_{y}(0)]}\label{eq:azxyz}
\end{equation}
for all $x$ and real constant $c$, and where the velocity of propagation is the constant $g_{0}$ moving along the $x-$axis. The direction of motion depends on the two solutions $g_{0}=g^{\pm}_{0}$. The solutions are both $g^{\pm}_{0}<0$ and the wave moves to the left along the $x-$axis, see Fig.~\ref{fig:g0pm}.

Again, this result is connected to studying the extraordinary wave, which the linearized approximation insufficiently described in our paper and some information is missing. Therefore the solutions do not contain the solutions which move to the right compared to the type I solutions. Further investigation with higher order approximation is required to see the full properties of the extraordinary wave.

The solution was obtained by integrating the coupled ordinary differential equations~(\ref{eq:fxt}) and (\ref{eq:dealpartiala}) together with the initial data $a_{z}(x)=A_{0}(x)$ for $t=0$.
The final solution also contains the particular solution which modifies the former by an additional constant $a^{f}_{z}(x)=A_{0}(x)-q_{0}/M$ for $t=0$.

\section{\label{sec:conclusion} Conclusion}
We have investigated the problem of nonlinear wave evolution in a quantum vacuum in the framework of Born--Infeld electrodynamics. We have been looking for a detailed theoretical description of the electromagnetic shock wave formation and its possible absence in a nonlinear quantum vacuum. We have investigated the two counter--propagating waves in the framework of Born--Infeld electrodynamics: a problem that describes the finite amplitude electromagnetic wave counter-propagating with respect to the crossed electromagnetic field for two linearly (Born) and nonlinearly polarized waves (Born--Infeld). This study has been motivated by our previous work on photon--photon scattering in vacuum \cite{KadlecovaKornBulanov2019, KadlecovaMine2019, BulanovSoliton} in the Heisenberg--Euler approximation; there we investigated the simpler problem of two linearly polarized waves.

For the linearly polarized waves, which correspond to the crossed field configuration (${\bf E}\cdot{\bf B}=0$, i.e. $\mathfrak{G}^2=0$), the Born--Infeld Lagrangian reduces to the Born Lagrangian as its special subcase. We have investigated the field equations of Born electrodynamics, which are identical to the equations for the Born--Infeld electrodynamics for the crossed field ${\bf E}\cdot{\bf B}=0$ (and hence referred to as Born).



We have solved the Born field equations analytically assuming the solution in the form of a simple wave. We added the small amplitude perturbations and linearized the coefficients to study the singularity formation. We have showed that the system of equations decoupled for the ordinary wave. The solutions have the form of a nonlinear wave without dispersion in the linear approximation.

 We have presented and analyzed the analytical solutions in the Born theory for the $+$ and $-$ solutions. These correspond to the counter--propagating waves case $-$ and co--propagating waves case $+$. We have presented the analytical formulae for both cases and have shown explicitly that the only solutions for both cases are exceptional waves.

We have analyzed the wave breaking in detail. For both cases, the wave steepening factor reduces to zero ($f'^{\pm}=0$), therefore only exceptional waves are the solutions. In the case $-$, it results in the exceptional wave propagating in the forward direction with constant phase velocity. In the case $+$, the resulting exceptional wave propagates in the backward direction with constant phase velocity. This exceptional wave is not a real physical contribution to the outgoing radiation from the photon--photon scattering process in Born electrodynamics.

We have shown explicitly that the only solutions for both cases are exceptional waves: the exceptional wave solutions in Born--Infeld electrodynamics do not turn into shocks in our approximation. Such result is in agreement with publications \cite{Boillat1972,BoillatI,BoillatII}, and the conjecture \cite{Brenier2004}, that the shocks are not allowed to form in the Born-Infeld electrodynamics. Since the conjecture was shown as false \cite{NevesSerre2005}, there is still possibility of their existence. Further analysis requires further investigation. It would be necessary to investigate which initial conditions are large enough to create the shock waves or use higher order approximation in the fields in our study to see whether that would enable the shocks to be created.

To investigate the problem with the Born--Infeld Lagrangian, we have extended our previous study of linearly polarized beams to the more general case of nonlinearly polarized beams (i.e. ${\bf E}\cdot{\bf B}\neq0$, where the term $\mathfrak{G}^2$ is non--vanishing). We have also investigated the extraordinary wave propagation. We have assumed the simplest generalization of our setup in order to solve the field equations and investigate the nonlinear wave evolution. 
We have added weak linear amplitude corrections and have linearized the coefficients to study the singularity formation. We have obtained a set of three equations for two variables ($a_{z}(x,t)$ and $b_{y}(x,t)$) together with a free given function $a_{y}(t)$. 
We have shown that the field equations for Born--Infeld decouple and can be solved, which is one of the main results of this paper. 

Further, we have analyzed the equations by the method of characteristics. This has enabled us to discuss the possible shock wave development and to analyze the direction of motion of the resulting waves. The set of equations consists of two types: the type I equation is the nonlinear wave equation and the type II equations are the nonlinear wave equations with non-zero right hand side. The equations contain information about the ordinary and extraordinary waves in the vacuum. The equation of type I corresponds to the ordinary wave, the type II equations with $a_{y}$ on the right hand side correspond to the development of the extraordinary wave.

Through the analysis of shock wave development by the method of characteristics we have found the following properties: 

In the type I equation we have shown that the requirement on the phase velocities to be constant (physically relevant) causes that the only relevant solutions of the equation are the exceptional waves. The nonlinear form of the resulting equation agrees with our results in Born theory and the Heisenberg--Euler approximation \cite{KadlecovaKornBulanov2019, KadlecovaMine2019}.  

The type II equation with the right hand side $\partial_{t}a_{y}$ also gives only exceptional waves as solutions. We have shown that the only shock wave which exists in the Born--Infeld is the one given as the initial condition and that it just propagates further.
The interpretation is that the source given by $a_{y}$ on the right hand side of equation~(\ref{eq:a222}) in a form of a function (\ref{eq:azxy}) is too weak to create a strong shock wave, therefore shock cannot be produced \cite{Whitham}.

We have analyzed and plotted the phase velocities and have identified their directions of propagation. The phase velocities originating from the first two equations in the set (\ref{eq:a2}) are the phase velocities $v^{\pm}_{1,2}$. 
The phase velocities $v^{\pm}_{1}$ and $v^{\pm}_{2}$ are plotted in figures ~\ref{fig:phase1pm} and ~\ref{fig:phase2pm}. The phase velocities $v^{\pm}_{1}$ correspond to the counter--propagating waves $-$ and the phase velocities $v^{\pm}_{2}$ correspond to the co--propagating waves $+$.  
The phase velocities $v^{\pm}_{3,4}$ originate from the first and the third equations in the set (\ref{eq:a2}). The phase velocities $v^{\pm}_{3}$ and $v^{\pm}_{4}$ are plotted in figures~\ref{fig:phase3pm} and ~\ref{fig:phase4pm}. The phase velocities $v^{\pm}_{3}$ and $v^{\pm}_{4}$ correspond to the co--propagating waves $+$ since their values are negative.  

We have analyzed the direction of propagation of the exceptional solutions.
The solution of the type I equation (\ref{eq:firsteq}), where the velocity of propagation is the constant $-1/\nu_{0}$, moves in the direction given by the two solutions $\nu^{\pm}_{0}$. We have observed that $\nu^{-}_{0}$ has positive values, therefore the wave moves to the right along the $x$ axis and corresponds to the counter--propagating case $-$ of the two beams. We have observed that $\nu^{+}_{0}$ has negative values, therefore the wave moves to the left along the $x$ axis and the solutions correspond to the co--propagating case $+$.

The phase velocities for the extraordinary wave give different phase velocities (\ref{eq:vph3}) than the ones for the ordinary wave (\ref{eq:vph111}). Which is inconsistent with the fact that the nonlinear Born--Infeld electrodynamics describes the electromagnetic fields in an isotropic medium with a polarization--independent refractive index \cite{RebhanTurk, Schrodinger2}, the extraordinary wave should move in the same direction as the ordinary wave with the same phase velocity.
 When looking at the plots of $v^{\pm}_{3}$ and $v^{\pm}_{4}$, the figures \ref{fig:phase3pm} and \ref{fig:phase4pm} show incomplete results with very low phase velocities only for the co--propagating case $+$. The physical phase velocities are missing. It is caused definitely by the insufficient precision of the linear approximation we used in our paper to study extraordinary wave. Linear approximation is enough to use for description of ordinary wave but not for extraordinary wave. In order to obtain also the physical phase velocities and the same phase velocites $v^{\pm}_{1,2}=v^{\pm}_{3,4}$, we would have to use approximation to higher orders.

The direction of motion of the type II solutions (\ref{eq:a222}) 
depends on the two solutions $\nu^{\pm}_{0}$. The motion is  then governed by the negative values of $g^{\pm}_{0}$ and thus the wave moves to the left along the $x-$axis. The absence of waves moving to right is also caused by the insufficient precision of linear approximation used for calculation of extraordinary wave in our work. Further investigation with higher order approximation is required to see the full properties of the extraordinary wave.

The solutions have the form of nonlinear waves without dispersion in the linear approximation; we have shown that the only physically relevant solutions are the exceptional waves which do not turn into shocks. The result is in agreement with the conjecture \cite{Brenier2004}, that the shocks are not allowed to form in the Born-Infeld electrodynamics. The conjecture was shown to be false \cite{NevesSerre2005}, therefore there is still a possibility of their existence which requires further investigation.

To summarize the solutions of type I and type II equations: the only physically relevant solutions are exceptional waves; wave steepening does not occur in either case. Upon choosing the physical phase velocities from all possible phase velocities, we have needed to find the constant ratio of $a_{y}/a_{z}$. We have shown that, in our solutions, such requirement is satisfied only for the exceptional waves. This also sets the free function $a_{y}$ to a specific expression, characteristic for each type of equation (I or II).

These theoretical studies, like the one in this paper, are important from the fundamental physics point of view. Heisenberg--Euler electrodynamics (QED) is considered to reflect the reality of our world better than the alternative, nonlinear Born--Infeld electrodynamics. In the classical vacuum, the classical Born--Infeld electrodynamics with point charges is well--defined and does not suffer from the UV--divergence problems that Heisenberg--Euler QED has in the quantum vacuum. The Born--Infeld quantisation does not exist at the moment and even if it were to, such quantization might not be UV--divergence--free and might not be able to explain all electromagnetic phenomena. 

The importance of this investigation is also demonstrated by the recent study on the photon--photon scattering experiment at LHC \cite{EllisQED}. The process of photon--photon scattering is the most promising process to study today in order to get some answers to fundamental questions addressed in particle theoretical physics, alongside with photon splitting, especially in looking for the numerical value of the free Born--Infeld parameter $b$. The measurement of parameter $b$ in the Born--Infeld theory will fulfill the long-standing need to determine the free Born--Infeld constant. 
 

\appendix
\onecolumngrid{
\section{\label{sec:CoefRadiation} The coefficients for the background field}
The whole Appendix is related to the Section~\ref{sec:AbsenceOfShockWavesBI} where we investigate the counter--propagating beams in Born--Infeld electrodynamics.

The coefficients $\alpha_{0}$, $\beta_{0}$, $\gamma_{0}$, $\delta_{0}$, $\epsilon_{0}$, $\zeta_{0}$, $\eta_{0}$, $\theta_{0}$, $\iota_{0}$, $\lambda_{0}$ in the set of equations~(\ref{eq:n1}, \ref{eq:n2}, \ref{eq:n3}) have the form:

The coefficients in the second equation (\ref{eq:n2}) are:
\begin{equation}
\alpha_{0}=1+\frac{E_{0}^2}{b^2}\frac{1}{C},\quad
\beta_{0}=\frac{1}{b^2}\frac{E^2_{0}}{C},\quad
\gamma_{0}= \left(1-\cfrac{1}{b^2}E^2_{0}\right)-\cfrac{E^2_{0}\left(1-\cfrac{1}{b^2}E^2_{0}\right)^2+\cfrac{1}{b^2}E^4_{0}}{b^2 C},\quad
\delta_{0}=\frac{1}{b^2}\cfrac{E^2_{0}}{C}.\label{eq:coeffs33}
\end{equation}
The coefficients in the third equation (\ref{eq:n3}) are:
\begin{align}
\epsilon_{0}&=\frac{1}{b^2}E^2_{0}\cfrac{\left(1+\cfrac{1}{b^2}E^2_{0}\right)}{C},\quad
\zeta_{0}=\frac{E^2_{0}}{b^2}\left(1-\cfrac{E^2_{0}\left(1-\cfrac{1}{b^2}E^2_{0}\right)}{b^2 C}\right),\quad
\eta_{0}= -\cfrac{E^2_{0}}{b^2}\left(2-\cfrac{\left(1-\cfrac{1}{b^2}E^2_{0}\right)\left(1+\cfrac{1}{b^2}E^2_{0}\right)}{C}\right),\label{eq:coeffs33}\\
\theta_{0}&=\frac{E^2_{0}}{b^2}\alpha_{0},\quad
\tau_{0}=1-\frac{E^2_{0}}{b^2},\label{eq:coeffs35}
\end{align}
where the constant $C$ reads
\begin{equation}
C=1-\frac{E^2_{0}}{b^2}-\frac{E^4_{0}}{b^4}.\label{eq:C}
\end{equation}

\section{\label{sec:Coef66} The coefficients for the Born--Infeld Lagrangian}
This Appendix shows the derivation of the coefficients, $\alpha$, $\beta$, $\gamma$, $\delta$, $\epsilon$, $\zeta$,  $\eta$,  $\theta$, $\iota$ and $\tau$, (\ref{eq:alphabeta}).

The $\alpha$ and $\beta$ coefficients, $\alpha_{a_{z}}, \alpha_{b_{y}}, \alpha_{a_{y}}$ and $\beta_{a_{z}}, \beta_{b_{y}}, \beta_{a_{y}}$ are:
\begin{align}
\alpha_{a_{z}}&=\frac{2E_{0}}{b^2C}\left(1+\frac{E^2_{0}}{b^2}\right),\;
\alpha_{b_{y}}=-\frac{2E^{3}_{0}}{b^4C^2}\left(1-\frac{E^2_{0}}{b^2}\right),\;
\alpha_{a_{y}}=\frac{2E^{3}_{0}}{b^4C^2}\left(1+\frac{E^2_{0}}{b^2}\right),\label{eq:koef144}\\
\beta_{a_{z}}&=\frac{E_{0}}{b^2C}\left(1+\frac{E^2_{0}}{b^2C}\right),\;\beta_{b_{y}}=\frac{E_{0}}{b^2C}\left[1-\frac{2E^2_{0}}{b^2C}\left(1-\frac{E^2_{0}}{b^2}\right)\right],\;
\beta_{a_{y}}=\frac{2E^3_{0}}{b^4C^2}\left(1+\frac{E^2_{0}}{b^2}\right).\nonumber
\end{align}
The $\gamma$ coefficients $\gamma_{a_{z}}, \gamma_{b_{y}}, \gamma_{a_{y}}$ are:
\begin{align}
\gamma_{a_{z}}&=-\frac{E_{0}}{b^4C}\left\{E^2_{0}+\frac{2E^2_{0}}{b^2C}\left[E^2_{0}+\left(1-\frac{E_{0}}{b^2}\right)^2\right]\right\},\nonumber\\
\gamma_{a_{y}}&=-\frac{2E_{0}}{b^2}-\frac{2E_{0}}{b^4C}\left[E^2_{0}-\frac{2E^2_{0}}{b^2}\left(1-\frac{E^2_{0}}{b^2}\right)+\left(1-\frac{E^2_{0}}{b^2}\right)^2\right]-\frac{2E^3_{0}}{b^6C^2}\left[E^2_{0}+\left(1-\frac{E^2_{0}}{b^2}\right)^2\right],\label{eq:koef14}\\
\gamma_{b_{y}}&=\frac{E^3_{0}}{b^4C}\left\{-1+\frac{2}{b^2C}\left(1-\frac{E^2_{0}}{b^2}\right)\left[E^2+\left(1-\frac{E^2_{0}}{b^2}\right)^2\right]\right\}.\nonumber
\end{align}
The $\delta$ coefficients $\delta_{a_{z}}, \delta_{b_{y}}, \delta_{a_{y}}$ are:
\begin{align}
\delta_{a_{z}}&=\frac{E_{0}}{b^2C}\left(1+\frac{2E^2_{0}}{C^2}\right),\;
\delta_{b_{y}}=-\frac{2E^3_{0}}{b^4C^2}\left(1-\frac{E^2_{0}}{b^2}\right),\;
\delta_{a_{y}}=\frac{E_{0}}{b^2C}\left[1+\frac{2E^2_{0}}{b^2C}\left(1+\frac{E^2_{0}}{b^2}\right)\right].\label{eq:koef145}
\end{align}
The $\epsilon$ coefficients  $\epsilon_{a_{z}}, \epsilon_{b_{y}}, \epsilon_{a_{y}}$ are:
\begin{align}
\epsilon_{a_{z}}&=\frac{E_{0}}{b^2C}\left(1+\frac{E^2_{0}}{b^2}\right)\left(1+\frac{2E^2_{0}}{b^2C}\right),\;
\epsilon_{a_{y}}=\frac{E_{0}}{b^2C}\left(1+\frac{E^2_{0}}{b^2}\right)\left[1+\frac{2E^2_{0}}{b^2C}\left(1+\frac{E^2_{0}}{b^2}\right)\right],\;
\epsilon_{b_{y}}=\frac{2E^7_{0}}{b^C}.\label{eq:koef15}
\end{align}
The $\zeta$ coefficients $\zeta_{a_{z}}, \zeta_{b_{y}}, \zeta_{a_{y}}$ are:
\begin{align}
\zeta_{a_{z}}&=\frac{E_{0}}{b^2}\left[1-\frac{E^2_{0}}{b^2C}\left(1+\frac{2E^2_{0}}{b^2C}\right)\right],\;
\zeta_{a_{y}}=\frac{E_{0}}{b^2}\left[1-\frac{E^2_{0}}{b^2C}\left(1-\frac{E^2_{0}}{b^2}\right)\left(1+\frac{2E^3_{0}}{b^4C}\left(1+\frac{E^2_{0}}{b^2}\right)\right)\right],\label{eq:koef154}\\
\zeta_{b_{y}}&=-\frac{E^3_{0}}{b^4C}\left[\frac{E^2_{0}}{b^2}-\left(1-\frac{E^2_{0}}{b^2}\right)+\frac{E^2_{0}}{b^2C}\left(1-\frac{E^2_{0}}{b^2}\right)^2\right].\nonumber
\end{align}
The $\eta$ coefficients $\eta_{a_{z}}, \eta_{b_{y}}, \eta_{a_{y}}$ are:
\begin{align}
\eta_{a_{z}}&=\frac{2E^3_{0}}{b^4C^2}\left(1-\frac{E^4_{0}}{b^4}\right),\;
\eta_{a_{y}}=\frac{2E^3_{0}}{b^4C}\left(1-\frac{E^2_{0}}{b^2}\right)\left[\frac{1}{C}\left(1+\frac{E^2_{0}}{b^2}\right)^2-1\right]-\frac{E_{0}}{b^2}\left[2-\frac{1}{C}\left(1-\frac{E^4_{0}}{b^4}\right)\right],\label{eq:koef155}\\
\eta_{b_{y}}&=-\frac{2E^3_{0}}{b^4C}\left(1+\frac{E^2_{0}}{b^2}\right)\left[\frac{1}{C}\left(1-\frac{E^2_{0}}{b^2}\right)^2+1\right]
-\frac{E_{0}}{b^2}\left[2-\frac{1}{C}\left(1-\frac{E^4_{0}}{b^4}\right)\right].\nonumber
\end{align}
The $\theta$ coefficients $\theta_{a_{z}}, \theta_{b_{y}}, \theta_{a_{y}}$ are:
\begin{align}
\theta_{a_{z}}&=\frac{2E^3_{0}}{b^2C}\left(1+\frac{E^2_{0}}{b^2C}\right),\;
\theta_{a_{y}}=\frac{E^3_{0}}{b^2C}\left[+\frac{2E^2_{0}}{b^2C}\left(1+\frac{E^2_{0}}{b^2}\right)\right]+E_{0},\;
\theta_{b_{y}}=\frac{E^3_{0}}{b^2C}\left[1-\frac{2E^2_{0}}{b^2C}\left(1-\frac{E^2_{0}}{b^2}\right)\right]+E_{0}.\label{eq:koef166}
\end{align}
The $\iota$ and $\tau$ coefficients, $\iota_{a_{z}}, \iota_{b_{y}}, \iota_{a_{y}}$ and $\tau_{a_{z}}, \tau_{b_{y}}, \tau_{a_{y}}$ are:
\begin{align}
\iota_{a_{z}}&=\frac{2E^3_{0}}{b^4C^2}\left(1-\frac{E^4_{0}}{b^4}\right),\;
\iota_{a_{y}}=\frac{2E_{0}}{b^2}+\frac{2E_{0}}{b^2C}\left(1-\frac{E^2_{0}}{b^2}\right)\left[1+2\frac{E^2_{0}}{b^2}+\frac{E^2_{0}}{b^2C}\left(1+\frac{E^2_{0}}{b^2}\right)^2\right],\label{eq:koef177}\\
\iota_{b_{y}}&=-\frac{2E^3_{0}}{b^4C}\left(1+\frac{E^2_{0}}{b^2}\right)\left[1+\frac{1}{C}\left(1-\frac{E^2_{0}}{b^2}\right)^2\right],\;
\tau_{a_{z}}=0,\,\tau_{a_{y}}=-\frac{2 E_{0}}{b^2},\,\tau_{b_{y}}=0.\label{eq:koef18}
\end{align}

\section{\label{sec:Coef23} The coefficients for the cases 1 and 2 of the total differential~(\ref{eq:56})}

We obtained two cases/solutions of the total differential $\left({\rm d}b_{y}/{\rm d}a_{z}\right)_{1,2}$, which have the form (\ref{eq:56}) together with expression(\ref{eq:78}), where
we have denoted the total differential as $\nu$ (\ref{eq:relation12}) and linearized it in the variables $a_{z}$, $b_{y}$, $a_{y}$ around a constant field using coefficients $\nu_{a_{z}}$, $\nu_{b_{y}}$, $\nu_{a_{y}}$. In what follows we will use the notation with $\pm$ used in equation (\ref{eq:nununu}) to distinguish between the two cases of solutions.

The $\nu_{\pm}$ coefficients are the following:
\begin{align}
\nu^{\pm}_{0}=\left(b^4B^2\left(2b^6E^6_{0}-b^4E^8_{0}-2b^2E^{10}_{0}+E^{12}_{0}+b^4E^4_{0}B+E^4_{0}B^2\pm B^3T\right)\right)/2B^3A,\label{eq:nupm}
\end{align}
where  we have denoted the larger expressions $A$, $B$, $D$, $T$ as
\begin{align}
A=&b^{12}-2b^{10}E^2_{0}-4b^8E^4_{0}+3b^{6}E^{6}_{0}+7b^4E^8_{0}-2b^2E^{10}_{0}-2E^{12}_{0},\nonumber\\
B=&b^4-b^2E^2_{0}-E^4_{0},\label{eq:params}\\
D=&-4b^{30}+12b^{28}E^2_{0}+36b^{26}E^4_{0}-104b^{24}E^{6}_{0}-152b^{22}E^8_{0}+352b^{20}E^{10}_{0}+429b^{18}E^{12}_{0}-618b^{16}E^{14}_{0}-743b^{14}E^{16}_{0}+536b^{12}E^{18}_{0}\nonumber\\
&+724b^{10}E^{20}_{0}-168b^{8}E^{22}_{0}-348b^{6}E^{24}_{0}-24b^{4}E^{26}_{0}+64b^{2}E^{28}_{0}+16E^{30}_{0},\nonumber\\
T=&\sqrt{(1/b^6B^6)D}.\nonumber
\end{align}

The coefficients $\nu_{a_{z}}^{\pm}$ are:
\begin{align}
\nu_{a_{z}}^{\pm}&=\left(\mp 4b^{42}E_{0} \mp 80b^2E_{0}^{41} \mp 16E^{43}_{0} \pm b^{24}E^{19}_{0}(3065 \mp 632T)\pm 2b^{16}E^{27}_{0}(1898 \mp 159T) \pm 2b^{32}E^{11}_{0}(354 \mp 61T)\right.\nonumber\label{eq:nuazpm}\\
&\pm \left.2b^{38}E^{5}_{0}(1 \mp 10T) \pm 16b^{6}E^{37}_{0}(27 \mp 2T)+2b^{36}E^{7}_{0}(\mp 95+T) \pm 8b^{4}E^{39}_{0}(\pm 3+T)+8b^{8}E^{35}_{0}(\pm 78+T)\right.\nonumber\\
&\left.+2b^{40}E^{3}_{0}(\pm 11+2T)+4b^{12}E^{31}_{0}(\mp 553+23T)+14b^{20}E^{23}_{0}(\mp 289+40T)+2b^{10}E^{33}_{0}(\mp 324+83T)
+b^{34}E^{9}_{0}(\pm 174+127T)\right.\nonumber\\
&\left.+b^{30}E^{13}_{0}(892 \pm 321T)+b^{14}E^{29}_{0}(432 \pm 361T)+b^{28}E^{15}_{0}(\mp 1725+416T)+b^{26}E^{17}_{0}(\pm 2282+453T)\right.\nonumber\\
&\left.+b^{18}E^{25}_{0}(\pm 2489+481T)+b^{22}E^{21}_{0}(3325 \pm 493T)\right)/2B^5A^2T.
\end{align}

The coefficients $\nu_{b_{y}}^{\pm}$ are:
\begin{align}
\nu_{b_{y}}^{\pm}&=\left(\mp 4b^{42}E_{0} \pm 72b^2E_{0}^{41} \pm 16E^{43}_{0} \pm b^{22}E^{21}_{0}(8689 \mp 1667T) \pm 2b^{14}E^{29}_{0}(6632 \mp 513T) \pm 2b^{24}E^{19}_{0}(3495 \mp 382T)\right.\nonumber\\
&\pm \left.2b^{30}E^{13}_{0}(1306 \mp 337T) \pm 2b^{16}E^{27}_{0}(1190 \mp 73T) \pm 4b^{12}E^{31}_{0}(40 \mp 21T) \pm 16b^{32}E^{11}_{0}(23 \mp 5T)+87b^{34}E^{9}_{0}(\mp 2+T)\right.\nonumber\\
&+\left.8b^{4}E^{31}_{0}(\pm 11+T)+2b^{40}E^{3}_{0}(\pm 11+2T)+2b^{10}E^{33}_{0}(\mp 1006+5T)+4b^{8}E^{35}_{0}(\mp 132+5T)
+4b^{6}E^{37}_{0}(\pm 9+7T)\right.\nonumber\\
&\left.+2b^{36}E^{7}_{0}(\mp 31+12T)+2b^{38}E^{5}_{0}(5 \pm 12T)+3b^{28}E^{15}_{0}(\mp 511+44T)+2b^{20}E^{23}_{0}(\mp 2211+245T)\right.\nonumber\\
&\left.+b^{26}E^{17}_{0}(\mp 4392+1063T)+b^{18}E^{25}_{0}(\mp 10091+1355T)\right)/2B^5A^2T.\label{eq:nubypm}
\end{align}

The coefficients $\nu_{a_{y}}^{\pm}$ are:
\begin{align}
\nu_{a_{y}}^{\pm}&=\left(\mp 2b^{44}E_{0} \pm 8b^{42}E_{0}^{3} \mp 92b^{4}E^{41}_{0} \mp 96b^{2}E^{43}_{0} \mp 16E^{45}_{0} \pm 2b^{14}E^{31}_{0}(45 \mp 46T)+8b^{38}E^{7}_{0}(\mp 7+T)\right.\nonumber\\
&+\left.b^{40}E^{5}_{0}(\pm 10+T)+8b^{6}E^{39}_{0}(\pm 60+T) \mp 2b^{10}E^{35}_{0}(384 \pm 7T)+2b^{8}E^{37}_{0}(\pm 497+17T)+b^{34}E^{11}_{0}(\pm 185+18T)\right.\nonumber\\
&+\left.19b^{26}E^{19}_{0}(\pm 55+37T)-6b^{12}E^{33}_{0}(\pm 511+37T) \mp b^{36}E^{9}_{0}(44 \pm 57T)+2b^{32}E^{13}_{0}(\pm 139+158T)
\mp b^{30}E^{15}_{0}(503 \pm 252T)\right.\nonumber\\
&\left.+2b^{16}E^{29}_{0}(\pm 2443+339T)+b^{18}E^{27}_{0}(\pm 1120+479T) \mp b^{28}E^{17}_{0}(1081 \pm 808T) \mp b^{22}E^{23}_{0}(1493 \pm 849T)\right.\nonumber\\
&\left. \mp b^{20}E^{25}_{0}(4662 \pm 1135T)+b^{24}E^{21}_{0}(\pm 2783+1190T)\right)/b^2B^5A^2T.\label{eq:nuaypm}
\end{align}

\section{\label{sec:Coef32} The coefficients for the function $g(a_{z},b_{y},a_{z})$}
The coefficients $g^{\pm}_{0}, g^{\pm}_{a_{z}}, g^{\pm}_{b_{y}}, g^{\pm}_{a_{y}}$, in the linearized form of the function $g(a_{z}, b_{y}, a_{y})$ (\ref{eq:gazayzz}), have the form:

\begin{align}\label{eq:g0}
g^{\pm}_{0}&=-\left\{\frac{1}{\alpha_{0}}(1+\tau_{0})\beta_{0}+\frac{\gamma_{0}}{\alpha_{0}}\nu^{\pm}_{0}\right\},
\end{align}
and 
\begin{align}\label{eq:qazqbyqay}
g^{\pm}_{a_{z}}&=\frac{\alpha_{a_{z}}}{\alpha^2_{0}}(\beta_{0}+\gamma_{0}\nu^{\pm}_{0}+\beta_{0}\tau_{0})-\frac{1}{\alpha_{0}}(\beta_{a_{z}}+\gamma_{a_{z}}\nu^{\pm}_{0}+\gamma_{0}\nu^{\pm}_{a_{z}}+\beta_{a_{z}}\tau_{0}+\beta_{0}\tau_{a_{z}}),\nonumber\\
g^{\pm}_{b_{y}}&=\frac{1}{\alpha^3_{0}}\left[-2\alpha_{a_{y}}\alpha_{a_{z}}(\beta_{0}+\gamma_{0}\nu^{\pm}_{0}+\beta_{0}\tau_{0})+\alpha_{0}\alpha_{a_{z}}(\beta_{a_{y}}+\gamma_{a_{y}}\nu^{\pm}_{0}+\gamma_{0}\nu^{\pm}_{a_{y}}+\beta_{a_{y}}\tau_{0}+\beta_{0}\tau_{a_{y}})\right.\nonumber\\
+&\left.\alpha_{0}\alpha_{a_{y}}(\beta_{a_{z}}+\gamma_{a_{z}}\nu^{\pm}_{0}+\gamma_{0}\nu^{\pm}_{a_{z}}+\beta_{a_{z}}\tau_{0}+\beta_{0}\tau_{a_{z}})-\alpha^2_{0}(\gamma_{a_{z}}\nu^{\pm}_{a_{y}}+\gamma_{a_{y}}\nu^{\pm}_{a_{z}}+\beta_{a_{z}}\tau_{a_{y}}+\beta_{a_{y}}\tau_{a_{z}})\right],\\
g^{\pm}_{a_{y}}&=\frac{1}{\alpha^4_{0}}\left\{\alpha_{0}\left(-2\alpha_{a_{y}}\alpha_{a_{z}}(\beta_{b_{y}}+\gamma_{b_{y}}\nu^{\pm}_{0}+\gamma_{0}\nu^{\pm}_{b_{y}}+\beta_{b_{y}}\tau_{0}+\beta_{0}\tau_{b_{y}})+ \alpha_{0}\alpha_{a_{z}}(\gamma_{a_{y}}\nu^{\pm}_{b_{y}}+\gamma_{b_{y}}\nu^{\pm}_{a_{y}}+\beta_{a_{y}}\tau_{b_{y}}+\beta_{b_{y}}\tau_{a_{y}})\right.\right.\nonumber\\
+&\left.\left.\alpha_{0}\alpha_{a_{y}}(\gamma_{a_{z}}\nu^{\pm}_{b_{y}}+\gamma_{b_{y}}\nu^{\pm}_{a_{z}}+\beta_{a_{z}}\tau_{b_{y}}+\beta_{b_{y}}\tau_{a_{z}})\right)\right.\nonumber\\
+&\left.\alpha_{b_{y}}\left(6\alpha_{a_{y}}\alpha_{a_{z}}(\beta_{0}+\gamma_{0}\nu^{\pm}_{0}+\beta_{0}\tau_{0})-2\alpha_{0}\alpha_{a_{y}}(\beta_{a_{z}}+\gamma_{a_{z}}\nu^{\pm}_{0}+\gamma_{0}\nu^{\pm}_{a_{z}}+\beta_{a_{z}}\tau_{0}+\beta_{0}\tau_{a_{z}})\right.\right.\nonumber\\
+&\left.\left.\alpha_{0}(-2\alpha_{a_{z}}(\beta_{a_{y}}+\gamma_{a_{y}}\nu^{\pm}_{0}+\gamma_{0}\nu^{\pm}_{a_{y}}+\beta_{a_{y}}\tau_{0}\tau_{a_{y}})+\alpha_{0}(\gamma_{a_{z}}\nu^{\pm}_{a_{y}}+\gamma_{a_{y}}\nu^{\pm}_{a_{z}}+\beta_{a_{z}}\tau_{a_{y}}+\beta_{a_{y}}\tau_{a_{z}}))\right)\right\}\nonumber.
\end{align}

\section{\label{sec:Coef44} The coefficients for the function $q(a_{z},b_{y},a_{z})$}
The coefficients $q_{0}, q_{a_{z}}, q_{b_{y}}, q_{a_{y}}$, in the linearized form of the function $q(a_{z},b_{y},a_{y})$ (\ref{eq:qqq}),
have the form:

\begin{align}\label{eq:q0}
q_{0}&=-\frac{\delta_{0}}{\alpha_{0}},
\end{align}
and 
\begin{align}\label{eq:qazay}
q_{a_{z}}&=\frac{1}{\alpha_{0}}\left(\frac{\alpha_{a_{z}}}{\alpha_{0}}\delta_{0}-\delta_{a_{z}}\right),\;
q_{b_{y}}=\frac{1}{\alpha_{0}^2}\left(-2\alpha_{b_{y}}\alpha_{a_{z}}\frac{\delta_{0}}{\alpha_{0}}+\alpha_{a_{z}}\delta_{b_{y}}+\alpha_{b_{y}}\delta_{a_{z}}\right)\,\nonumber\\
q_{a_{y}}&=\frac{1}{\alpha^{3}_{0}}\left[2\alpha_{b_{y}}\alpha_{a_{y}}\left(3\alpha_{a_{z}}\frac{\delta_{0}}{\alpha_{0}}-\delta_{a_{z}}\right)-2\alpha_{a_{z}}(\alpha_{a_{y}}\delta_{b_{y}}+\alpha_{b_{y}}\delta_{a_{y}})
\right].
\end{align}

\begin{acknowledgments}

H. K. wishes to thank: Prof.~I.~Bia{\l}ynicki--Birula for enlightening and kind discussions on Born and the Born--Infeld theory, their relativistic covariance, and his pointing my attention to the beauty of the original Born--Infeld paper;
Dr.~T.~Pech\'{a}\v{c}ek for discussions on relativistic covariance and his detailed final reading of the paper; Prof.~S.~Bulanov for discussions and his interest in the manuscript; Dr.~T.~Chrobok for discussion and his interest in this work; Dr. Ch. Lu for helful discussions on the submission process and his interest in this work; Prof. G. Gibbons for a discussion about plane wave solutions; Prof.~G.~Gregori for pointing our attention to the corrected version of the paper about PVLAS experiment; Dr.~E. Chacon--Golcher for a very detailed and thorough final reading of the manuscript and his comments.

H.K. is grateful for kind, supportive and helpful report from the anonymous referee who pointed my attention to the mathematical work on shock wave development by prof. D. Christodoulou, Y. Brenier and D. Serre, which I was not aware of and which motivated further investigation.

The work was supported by the project High Field Initiative (CZ$.02.1.01/0.0/0.0/15\_003/0000449$) from European Regional Development Fund. H. K. was supported by the fellowship (award) Czech edition of L'Or\'{e}al UNESCO For Women In Science 2019.  
\end{acknowledgments}
\twocolumngrid
\bibliography{apssampELI12}

\end{document}